\documentclass[reprint,superscriptaddress,amsmath,amssymb,aps,prb,floatfix]{revtex4-2}

\usepackage[utf8]{inputenc}
\DeclareUnicodeCharacter{0301}{\'{}}
\usepackage[normalem]{ulem} 
\usepackage[acronym,toc,automake=immediate,nonumberlist,nopostdot,nogroupskip]{glossaries}

\usepackage[hidelinks]{hyperref}
\hypersetup{
    colorlinks=true,
    breaklinks=true,
    linkcolor=blue,
    filecolor=magenta,      
    urlcolor=blue,
    pdftitle={Review copy --- Magnetic interactions and spin orders in \texorpdfstring{Cr$_8$}{Cr8} and \texorpdfstring{V$_8$}{V8} ring-shaped molecular magnets from non-collinear ab initio calculations},
    pdfpagemode=FullScreen,
}

\usepackage{mathtools,color}
\usepackage{braket}
\usepackage{chemformula} 

\let\ce\ch

\usepackage{subfigure}
\usepackage{graphicx}

\usepackage{xspace}
\usepackage{multirow}

\usepackage{etoolbox}

\AtBeginDocument{
  
  \patchcmd{\bibinfo}{\doibase}{}{}{}
  \patchcmd{\bibinfo}{\url}{\relax}{}{}{}
  \patchcmd{\bibinfo}{title}{dummytitle}{}{}
  \patchcmd{\bibinfo}{\textit}{\relax}{}{}{}
  \patchcmd{\bibfield}{title}{dummytitle}{}{}
  \patchcmd{\bibinfo}{howpublished}{dummyhp}{}{}
  \patchcmd{\bibinfo}{note}{dummynote}{}{}
  \patchcmd{\bibinfo}{eprint}{dummyeprint}{}{} 
}

\usepackage{tikz}
\usetikzlibrary{positioning}

\newcommand{\editor}[2]{%
  \expandafter\newcommand\csname #1note\endcsname[1]{%
    \textcolor{#2}{(\textbf{#1:} ##1)}}%
  \expandafter\newcommand\csname #1\endcsname[1]{%
    \textcolor{#2}{##1}}%
  \expandafter\newcommand\csname #1cancel\endcsname[1]{%
    \textcolor{#2}{\sout{##1}}}%
  \expandafter\newcommand\csname #1change\endcsname[2]{%
    \textcolor{#2}{\sout{##1} ##2}}%
  \newenvironment{#1text}{\color{#2}}{\color{black}}
}

\definecolor{red}{rgb}{1.00,0.00,0.00}
\definecolor{RED}{rgb}{1.00,0.00,0.00}
\definecolor{green}{rgb}{0.00,1.00,0.00}
\definecolor{blue}{rgb}{0.00,0.00,1.00} 
\definecolor{orange}{rgb}{1.00,0.50,0.00}
\definecolor{ORANGE}{rgb}{1.00,0.50,0.00}
\definecolor{magenta}{rgb}{1.00,0.00,1.00}
\definecolor{cyan}{rgb}{0.00,1.00,1.00}
\definecolor{brown}{rgb}{0.4,0.2,0.00}
\definecolor{deepsky}{rgb}{0.0,0.75,1.00}
\definecolor{gray}{rgb}{0.50,0.50,0.50}
\definecolor{navy}{rgb}{0.137,0.137,0.557}
\definecolor{burntorange}{rgb}{0.8, 0.33, 0.0}
\definecolor{asparagus}{rgb}{0.53, 0.66, 0.42}
\definecolor{emerald}{rgb}{0, 0.605, 0.465}
\definecolor{amethyst}{rgb}{0.6, 0.4, 0.8}
\definecolor{purple}{rgb}{0.502, 0, 0.502}
\definecolor{lava}{rgb}{0.81, 0.06, 0.13}
\definecolor{tk}{rgb}{0.5, 0.00, 0.8}
\definecolor{beaver}{rgb}{0.62, 0.51, 0.44}
\definecolor{yellow}{rgb}{0.7, 0.7, 0.0}
\definecolor{seagreen}{rgb}{0.18,0.55,0.34}
\definecolor{oxford}{rgb}{0.0,0.129,0.278}
\definecolor{lightgray}{rgb}{.9,.9,.9}
\definecolor{lightblue}{rgb}{0.0,0.0,0.8}
\definecolor{fuchsia}{rgb}{1.0,0.47,1.0}
\editor{LB}{magenta}
\editor{AF}{red}

\makeglossaries 

\newacronym{DFT}{DFT}{Density Functional Theory} 
\newacronym{TM}{TM}{transition metal} 
\newacronym{QE}{QE}{Quantum ESPRESSO} 
\newacronym{bs}{BS-DFT}{Broken-Symmetry DFT} 
\newacronym{BQ}{BQ}{biquadratic} 
\newacronym{DM}{DM}{Dzyaloshinskii-Moriya} 
\newacronym{AFM}{AFM}{antiferromagnetic} 
\newacronym{FM}{FM}{ferromagnetic} 
\newacronym{LDA}{LDA}{Local Density Approximation} 
\newacronym{xc}{xc}{exchange and correlation} 
\newacronym{soc}{SOC}{spin-orbit coupling} 

\def\restrict#1{\raise-.5ex\hbox{\ensuremath|}_{#1}}



\begin{document}

\title{
Magnetic interactions and spin ordering in \texorpdfstring{Cr$_8$}{Cr8} and \texorpdfstring{V$_8$}{V8} molecular magnets from non-collinear ab initio calculations}

\author{Maria Barbara Maccioni}
\affiliation{Department of Physics, University of Pavia, Via A. Bassi 6, I-27100 Pavia, Italy}

\author{Elia Stocco}
\affiliation{Max Planck Institute for the Structure and Dynamics of Matter, Hamburg, Germany}

\author{Luca Binci}
\altaffiliation{Present address: Department of Materials Science and Engineering, University of California Berkeley, Berkeley, California, USA.}
\affiliation{Theory and Simulation of Materials (THEOS), and National Centre for Computational Design and Discovery of Novel Materials (MARVEL), École Polytechnique Fédérale de Lausanne, CH-1015, Lausanne, Switzerland.}

\author{Andrea Floris}
\email[e-mail:]{afloris@lincoln.ac.uk, an.floris@gmail.com} 
\affiliation{Department of Chemistry, School of Natural Sciences, University of Lincoln, Brayford Pool, Lincoln LN6 7TS, United Kingdom}

\author{Matteo Cococcioni}
\email[e-mail:]{ matteo.cococcioni@unipv.it}
\affiliation{Department of Physics, University of Pavia, Via A. Bassi 6, I-27100 Pavia, Italy}

\date{\today}

\begin{abstract}

\noindent%

We investigate the magnetic interactions in the octanuclear molecular rings \ce{Cr8} and \ce{V8} using density functional theory within a non-collinear framework. By probing magnetic configurations with fully unconstrained three-dimensional spin orientations, we systematically explore the rotational degrees of freedom of neighboring spins and go beyond the conventional low-energy Heisenberg description. This approach distinguishes in-plane from out-of-plane exchange interactions, quantifies biquadratic spin couplings, and reveals significant antisymmetric exchange of the Dzyaloshinskii–Moriya type. The comparison with linear chains of equal composition reveals a clear dependence of the Dzyaloshinskii–Moriya interactions on the curvature of the annular structure.
Exact diagonalization of the spin Hamiltonian constructed from the computed effective couplings demonstrates the role of these unconventional interactions and identifies an exotic ground state in \ce{V8}, driven by the competition between nearest- and next-nearest-neighbor exchange couplings of opposite sign.
The closest quantitative agreement with experiment, particularly for \ce{Cr8}, is achieved using extended Hubbard functionals that explicitly incorporate intersite hybridization couplings.

\end{abstract}

\glsresetall

\maketitle

\color{black}

\section{Introduction}\label{sec:intro}

Magnetic molecular rings are a class of molecular nanomagnets that provide an ideal platform for exploring collective phenomena in low-dimensional systems emerging from microscopic magnetic interactions \cite{MolecularNanomagnets,Blachowicz}.
For more than two decades, ring-shaped molecular clusters such as \ce{Fe6}, \ce{Fe10} \cite{Cornia}, \ce{Cr8} \cite{affronte03, whinpenny,Waldmann03,van2002magnetic}, and \ce{Cu8} \cite{Ardizzoia} have been extensively investigated through techniques including nuclear magnetic resonance, specific heat measurements, and inelastic neutron scattering. These studies have consistently revealed \gls{AFM} exchange interactions between nearest-neighbor spins.

Within this broad class, Cr-based molecular rings, particularly \ce{Cr8}, have emerged as important model systems, motivating extensive experimental \cite{affronte03,whinpenny,Waldmann03,van2002magnetic} and theoretical \cite{bellini2010density,Kronik,chiesa16} studies because of their potential use in quantum information technologies \cite{chiesa23,chiesa14,chiesa20,ghirri17,timco09,timco16,candini10}.

In contrast, V-based rings remain relatively unexplored; the inherent instability of octanuclear vanadium rings, such as \ce{V8}, has limited the availability of well-characterized systems and hindered systematic investigations of their magnetic properties \cite{sorolla19, laye2003solvothermal}.
For this reason, research efforts have mostly focused on heterometallic magnetic wheels of the form V$_7$X, where X denotes a transition metal distinct from vanadium \cite{Lascialfari}. These heterometallic assemblies exhibit enhanced magnetic properties and structural robustness, making them more amenable to detailed characterization and functional studies. The incorporation of a foreign metal center often leads to novel magnetic behaviors -- such as nonvanishing total magnetization in doped \gls{AFM} systems -- making these materials attractive for applications based on molecular magnetism.

Theoretical modeling of such systems is essential for elucidating their quantum behavior. Key aspects of interest include the influence of geometrical curvature on spin ordering and magnetic exchange \cite{gentile13,gentile15}, the emergence of generalized \glspl{soc} \cite{cardias2020dzyaloshinskii,Cardias2020}, and the interplay between spin and transport properties \cite{PhysRevLett.102.246801}. Understanding these effects provides a microscopic basis for controlling magnetic and spin-dependent properties, with implications for quantum information processing, molecular spintronics, and magnetic refrigeration. 

In a recent work \cite{stocco2025}, we performed a comparative study of two ring-shaped molecular magnets based on Cr and V centers using collinear spin \gls{DFT}, with the aim of determining the effective exchange couplings between magnetic moments localized on the \gls{TM} centers.
Consistent with other molecular rings of similar structure, \ce{Cr8} exhibits an \gls{AFM} ground state dominated by nearest-neighbor exchange interactions. In contrast, \ce{V8} adopts a \gls{FM} ground state, with dominant negative (i.e., \gls{FM}) nearest-neighbor couplings partially counterbalanced by next-nearest-neighbor interactions of opposite sign (\gls{AFM}).
Through a series of targeted calculations with modified ligand groups around the \gls{TM} magnetic centers, we isolated the role of the metal oxidation state in determining the effective magnetic couplings and traced the anomalous behavior of \ce{V8} to its less-than-half-filled $d$-state submanifold.

Here, we extend 
the investigation of the two systems within a \textit{non-collinear} \gls{DFT} framework. This approach enables us to explore a broader space of magnetic configurations, including non-parallel moments with fully unconstrained, site-specific orientations. We use these configurations to characterize more general effective interactions between localized spins, including \gls{BQ} and \gls{DM} couplings, and to determine how these interactions depend on the structural curvature of the molecular rings.
To isolate the role of curved geometry, we compare the results for the annular molecules with those for a linear, periodic chain of equal composition and local bonding structure. This comparison provides a direct measure of the magnetic effects associated with the ring topology.
We further elucidate the role of the different interaction terms through exact diagonalization of the quantum spin-model Hamiltonian -- formulated as a generalized Heisenberg model -- by introducing the computed couplings one at a time. This analysis identifies the contributions of the individual interactions to the ground state and reveals their influence on the low-temperature magnetic behavior of the systems.

The work is organized as follows: in Section \ref{sec:02} we review the molecular structures in their annular and linear conformation, used for comparison. In Section \ref{sec:03} we outline the computational details of our calculations. Section \ref{sec:res} presents the main results, 
i.e., the calculation of the exchange couplings (Sec.~\ref{excoup}), of the biquadratic and Dzyaloshinskii-Moriya-type interactions (Secs.~\ref{bqt} and ~\ref{sec:DM}), and of the magnetic susceptibility and correlation functions  
(Sec.~\ref{sec:MP}).
Section \ref{sec:disc} 
summarizes and discusses the main findings and proposes directions for future research. 
In-depth analyses of particularly important aspects are finally presented in the appendices. 

\section{\label{sec:02}Ring structures and linear chain model}

\begin{figure}
\subfigure[Polyhedral structure of an isolated molecular \ce{M8}-ring. ]{\label{fig:fig1}
\includegraphics[width=70mm]{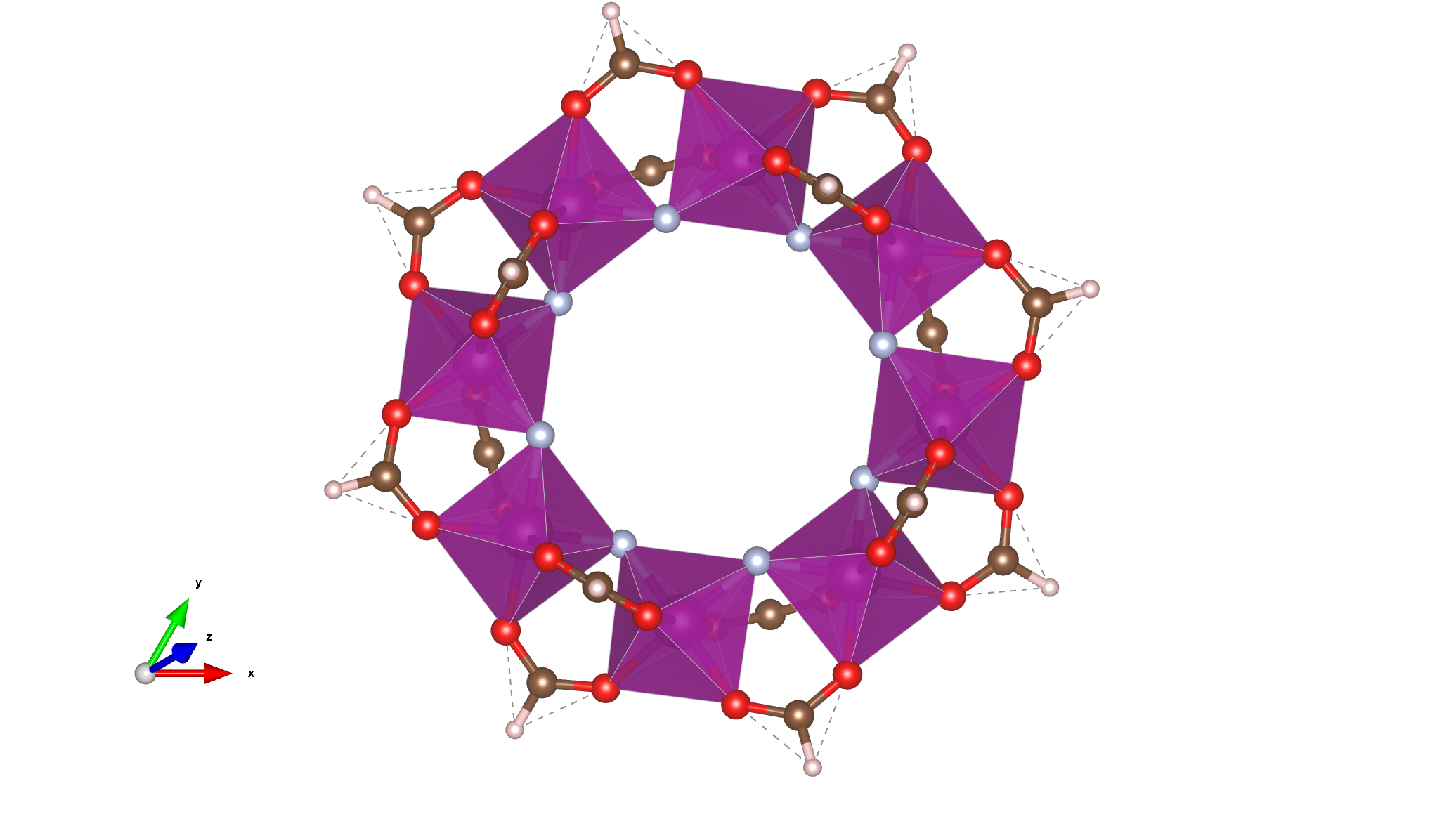}}
\subfigure[Unit cell reference for an octagonal heterometallic ring.]{
\label{fig:fig1-a}
\includegraphics[width=70mm]{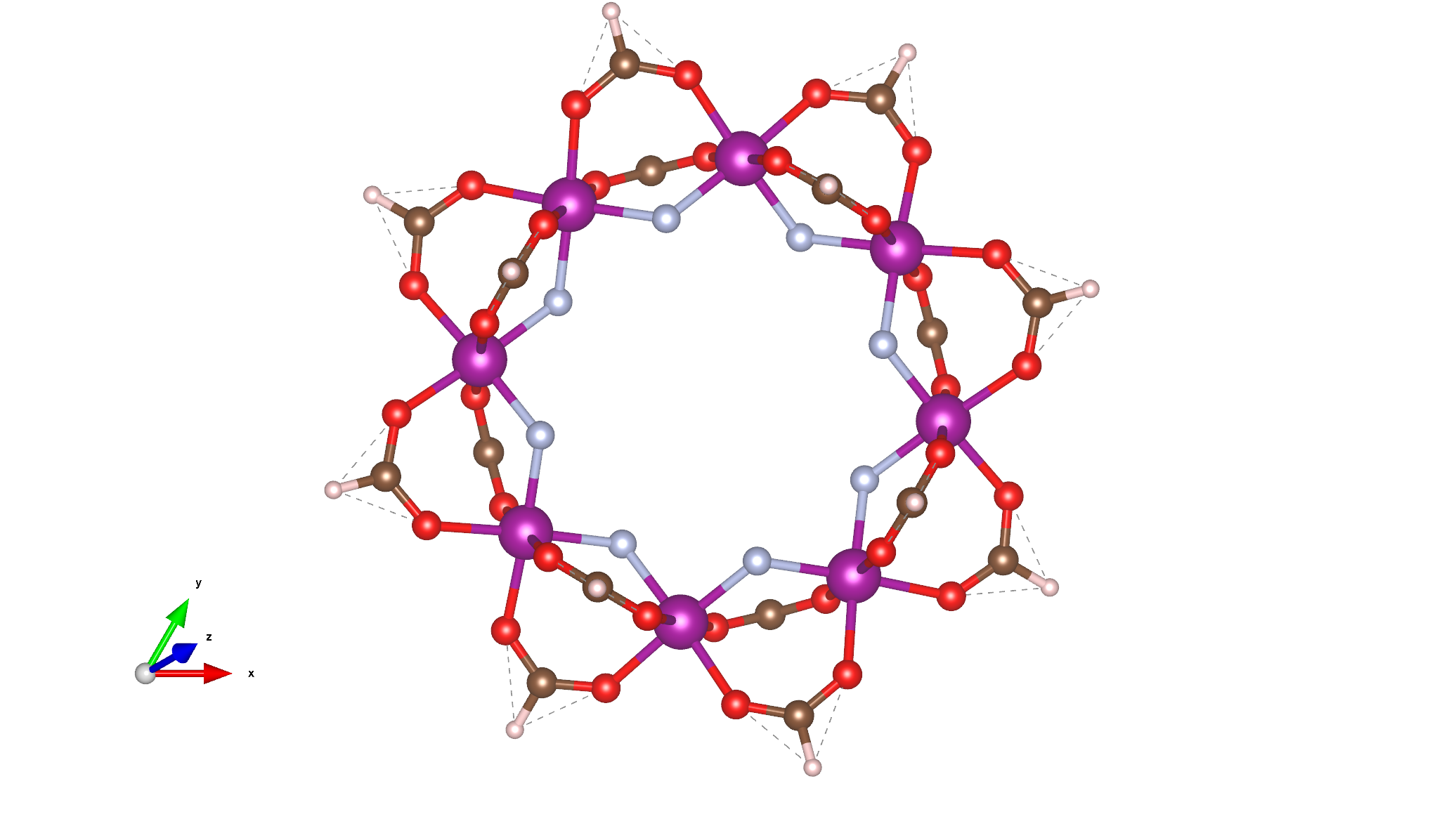}}
\centering
\caption{\ce{M8}-ring  structure reduced through the hydrogen saturation. Color code: M violet, F grey, O red, C brown, H white.}
\end{figure}

\begin{figure}
      \subfigure[Unit cell of the linear chain model.]{\label{fig:fig2}\includegraphics[width=50mm]{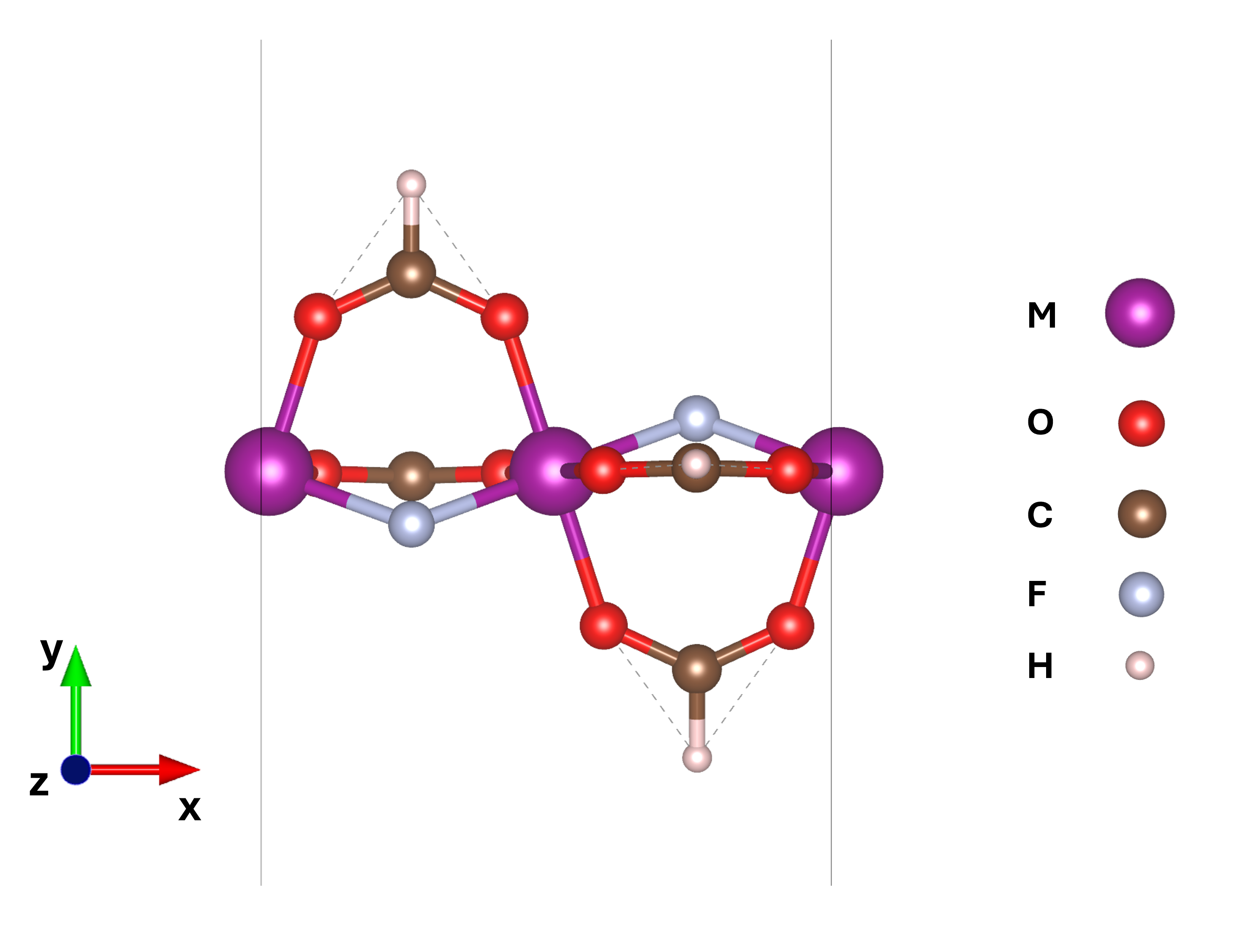}}
      \subfigure[Polyhedral structure of the chain model, illustrating the periodic arrangement along the $x$ axis.]{\label{fig:fig2-a}\includegraphics[width=60mm]{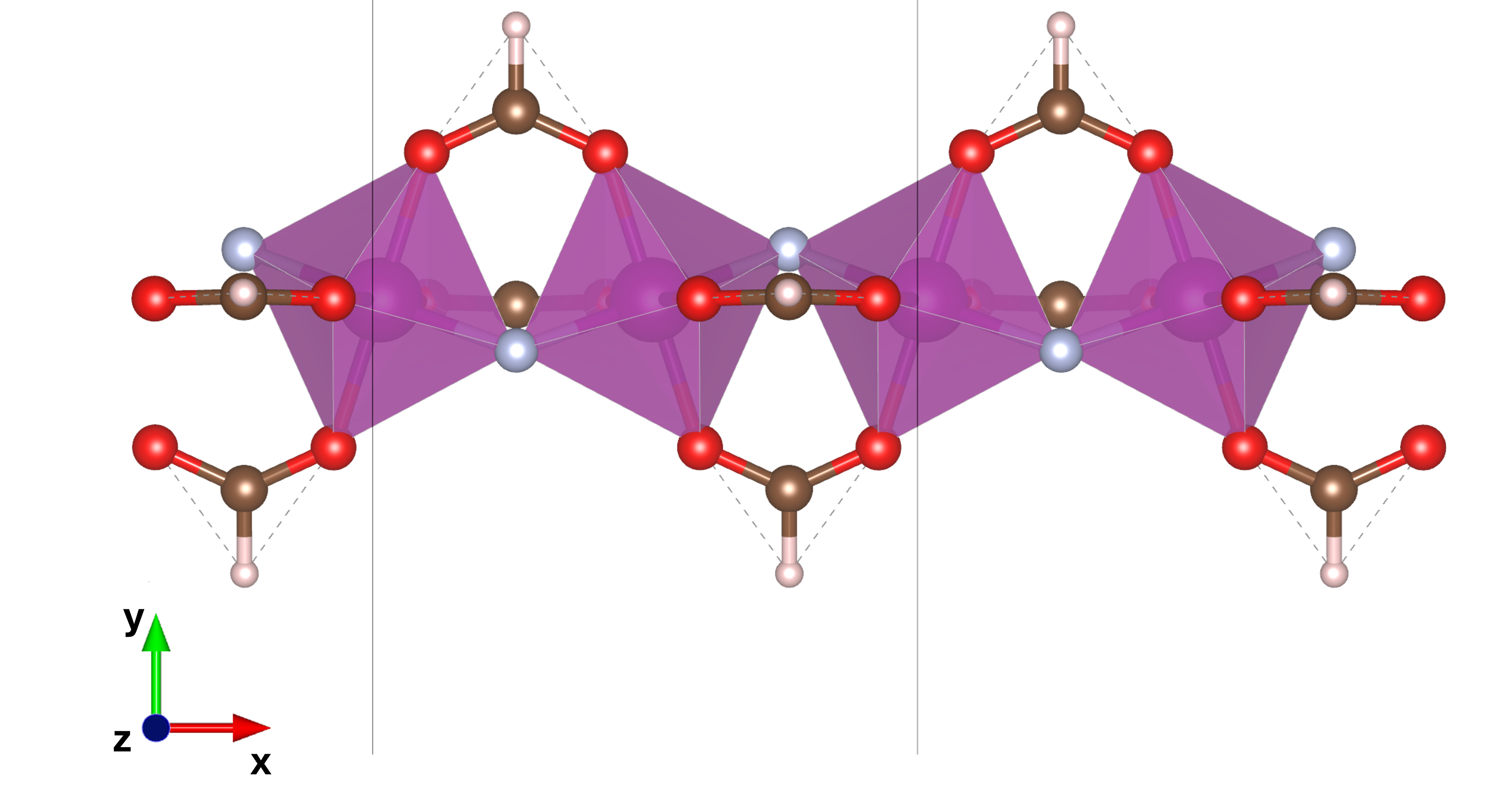}}
      \centering
      \caption{Representation of the M-chain molecular structures, unit-cell and supercell, adopted in this study. }
\end{figure}

The isolated octagonal heterometallic ring consists of eight almost coplanar \gls{TM} cations (the M$^{+3}$ magnetic centers) in octahedral coordination with surrounding oxygen and fluorine anions. These units share the fluorine corner on the inner space of the ring, as evident from Fig. \ref{fig:fig1}. On the outside they are connected to each other via two independent carboxylate --O--C--O-- bridges, fluorine bridge, and organic pivalic groups (see also Fig. \ref{fig:fig2}). These components form an almost perfectly planar octagonal structure \cite{van2002magnetic, PhysRevB.67.094405}. The compound adopts the formula unit \ce{M8F8Piv16}, where Piv refers to pivalic acid (trimethyl acetic acid). To simplify the organic matrix for computational modeling, the pivalate groups are replaced by hydrogen atoms, a procedure known as \textit{hydrogen saturation}. This results in a reduced molecular formula \ce{M8F8O2CH16}, corresponding to the structure shown in Fig. \ref{fig:fig1}, and \ref{fig:fig1-a}, and containing only 80 inequivalent atoms.

As previously mentioned, we also adopt a linear chain model of the considered systems, characterized by the same composition and local bonding structure (i.e., the same \gls{TM}-centered octahedra) as the annular molecules. A two-\gls{TM} linear chain of 20 atoms (Fig. \ref{fig:fig2}) was also adopted in our previous study \cite{stocco2025} in order to reduce the computational costs associated with the calculation of the Hubbard parameters. Here we also consider a four-\gls{TM} linear chain system of 40 atoms (see Fig. \ref{fig:fig2-a}) that allows us to consider more articulated spin configurations (helical, in particular) and to perform a thorough comparison with ring-shaped systems. This comparative analysis proves crucial to gain insight on how the geometric curvature of the molecule influences its magnetic behavior and, in particular, on its role in the emergence of chiral \gls{DM} and \gls{BQ} couplings. 

Following the work of Tomecka and collaborators \cite{Bellini2008}, we have accommodated our chain structure, illustrated in Fig.\ref{fig:fig2}, in an orthorhombic cell whose $x$ axis is the line that goes through the M atoms, and the $y$ and $z$ axes are, respectively, perpendicular and parallel to the plane defined by the original \ce{M8}-ring. The unit cell is periodically repeated along the $x$ axis, with the corresponding lattice parameter set to twice the M--M distance to ensure adequate separation between magnetic centers. The remaining lattice parameters are chosen to provide sufficient vacuum space between periodic replicas, thereby preventing any unphysical interaction. 
As mentioned previously the chain model maintains the same local structure of the corresponding ring-shaped molecule, with the same bond lengths and angles between the metal centers and the bridging atoms. However, over longer distances the sequence of the bridges differs; in particular, the alternation of F vertices on either side of the chain (consequent to a rigid rotation of some of the octahedral units around the $x$ axis) is necessary to make the system assume a linear structure.  

\section{\label{sec:03}Computational details}

Structural optimizations and self-consistent calculations were performed using the plane-wave pseudopotential implementation of \gls{DFT} \cite{DFT-HK,KS65} contained in the \gls{QE} 
package \cite{QE-2009,QE2017}.  
The exchange-correlation (xc) energy was approximated using the \gls{LDA} \cite{Alder,Gross,Parr}, while the interaction of valence electrons with the atomic cores was described by pseudopotentials based on the projector augmented-wave methodology \cite{PAW,PAW-dalCorso,DALCORSO2014337}. The electronic wavefunctions and charge density were expanded up to kinetic energy cutoffs of 80 Ry and 640 Ry, respectively. The ring molecules were treated as isolated systems at the center of empty simulation boxes able to guarantee a minimum spacing between periodic replicas of about 15 \AA. Correspondingly, the Brillouin zone was sampled through the $\Gamma$ point only.
 
The k-point mesh in the periodic chain models is chosen to facilitate a meaningful comparison with the ring structures. Assuming that a $\Gamma$-point sampling is sufficient for a linear chain of 8 \gls{TM}-centered units (same length as the ring), periodic along $x$, for the 2- and 4-unit chains we employed the $4\!\times1\!\times1$ and $2\!\times1\!\times1$ K-point meshes, respectively. 

In order to improve the description of electronic localization -- central to capture localized magnetic moments -- we used the \gls{DFT}+U approach that entails corrections to approximate \gls{xc} functionals, modeled on the Hubbard Hamiltonian \cite{raccolta,Himmetoglu,LRHubbard}. An extended version of \gls{DFT}+U, including inter-site interactions and denoted \gls{DFT}+U+V \cite{DFTUV}, was also employed, following its recent extension to spin non-collinear \gls{DFT} implementation in \gls{QE} \cite{LucaBinci}.

The effective Hubbard parameters $U$ and $V$ were computed using the linear-response method introduced in Ref. \cite{LRHubbard} through its density-functional-perturbation-theory implementation contained in the HP code of \gls{QE} \cite{Cococcioni2018,TIMROV2022108455}. These calculations were performed in a self-consistent fashion (on  the systems' reference geometry) as explained in Ref. \cite{cococcioni21}. For computational efficiency, the calculation of the Hubbard parameters was performed using the two-unit linear chain. As was previously reported \cite{stocco2025}, maintaining the same local bonding structure guarantees that the results are effectively the same as in the ring-shaped molecules.
The computed values are presented in Table \ref{table:Hubbard}. 

\begin{table}[h]
  \begin{tabular}{c|c r|c r } 
    \hline
        &\multicolumn{2}{c|}{\ce{Cr} } 	
        &\multicolumn{2}{c}{\ce{V} } \\
        & U(eV) & V(eV) & U(eV) & V(eV) \\
    \hline
    \multirow{3}{3em}{LDA-PAW}& 5.26	& Cr-Cr 6.28	& 4.62	& V-V 5.66	\\
    & 		& Cr-F 1.54	& 		& V-F 1.48	\\
    &		& Cr-O 1.40	&		& V-O 1.37  \\
    \hline
  \end{tabular}
 \caption{Calculated Hubbard parameters using \gls{LDA}.}
  \label{table:Hubbard}
\end{table}

The use of the extended \gls{DFT}+U+V approach leads to a significant ($\approx$ 1 eV) increase of the on-site Hubbard $U$ (larger for Cr than for V) while the $V$ between the \gls{TM} centers and nearest neighbor anions is about 1.4 - 1.5 eV.
\\

The method adopted here to evaluate the coupling parameters of a model spin Hamiltonian from \gls{DFT} calculations, is based on the non-collinear generalization of the \gls{bs} method, as presented in Ref. \cite{bsdft}. Given a spin configuration converged with \gls{DFT}, the magnetic moments $\mathbf{m}_i$ of the \gls{TM} ions are evaluated as $\mathbf{m}_i=g\mu_\mathrm{B}\mathbf{S}_i$ where $\mu_\mathrm{B}$ is the Bohr magneton, $g$ is the spin gyromagnetic factor (assumed equal to 2 in this work) and  $\mathbf{S}_i$ is the expectation value  of the spin of the site $i$ (typically obtained by integrating the magnetization density over a sphere centered on the same ion). Within spin non-collinear \gls{DFT} these moments point along directions that can be different on different sites; they can thus be assimilated to classical vectors obtained from the expectation value of the spin along site-specific quantization axes $\mathbf{n}_i=\mathbf{m}_i/|\mathbf{m}_i|$. The spin state of the whole system is represented by the tensor product of all the single-spin states $\Psi_S = \psi_1 \otimes \psi_2 \otimes \dots \otimes \psi_8$. The \gls{bs} method consists in matching \gls{DFT} total energies E$\left[\{\mathbf{S}_i\}\right]$ with the expectation value of the model spin Hamiltonian on a state $\Psi_S$ such that $\langle \Psi_S \vert \hat{\mathbf{\sigma}}_{\mathbf{n}_i} \vert \Psi_S\rangle = \mathbf{S}_i$. 

Careful but simple algebraic manipulation, whose results (without derivation) will be used in the next Section, can lead to analytical expressions for these expectation values in terms of the localized magnetic moments $\mathbf{S}_i$ and the angles between them.

Finally, the diagonalization of the spin Hamiltonian and the calculation of the magnetic susceptibility $\chi$,  presented in Section~\ref{sec:MP}, 
were performed using the custom Python code \texttt{QuantumSparse}, developed and made publicly available by one of the authors \cite{qs}.

\section{Results}\label{sec:res}

      \subsection{Exchange \texorpdfstring{$J$}{J} coupling: spin non-collinear DFT}\label{excoup}
      The use of non-collinear spin \gls{DFT} and the consequent re-definition of local spins as vectorial quantities (whose components are the expectation values of the corresponding Pauli matrices) imposes a generalization of the model Hamiltonian used to study low energy spin-excitations and to evaluate the effective spin-spin interactions. At the lowest level this generalization entails rewriting the interaction between localized spins on neighbor sites as proportional to their scalar product:

      \begin{equation}
          \hat H_\mathrm{ex} \equiv J \sum_i \hat{\mathbf S}_i \cdot \hat{\mathbf S}_{i+1}. 
          \label{Hex}
      \end{equation}

      The one given in Eq. \eqref{Hex} corresponds to an isotropic Heisenberg Hamiltonian, where the effective couplings $J$ (that are independent on the index $i$ due to the substantial equivalence of all the magnetic sites) are the same in all the Cartesian directions. A straightforward generalization of this expression can be obtained through the following expression:

      \begin{equation}
          \hat H_\mathrm{ex} \equiv \sum_i \hat{\mathbf S}_i \cdot {\mathbf J} \cdot \hat{\mathbf S}_{i+1}. 
          \label{Hex1}
      \end{equation}
      where the effective couplings are now included in a 3$\times$3 interaction matrix whose component $J^{\alpha\beta}$ couples $\hat{\mathbf S}_{i}^{\alpha}$ with $\hat{\mathbf S}_{i+1}^{\beta}$. Assuming that the off-diagonal elements of the interaction matrix ${\mathbf J}$ 
can be neglected one can easily recover the so-called XYZ model Hamiltonian: 
      \begin{align}
          \hat H_\mathrm{ex} =  & \,
            \sum_i \left[
              J^x \hat S_i^x \hat S_{i+1}^x + J^y \hat S_i^y \hat S_{i+1}^y + J^z \hat S_i^z \hat S_{i+1}^z
              \right]. 
              \label{heis2}
      \end{align}

      If $z$ is taken as the reference direction (e.g., the easy axis of magnetization, whenever present) and the couplings are assumed to be equal on the transverse plane, $J^x = J^y = J^z \Delta$, the Hamiltonian in Eq. \eqref{heis2} reduces to the isotropic Heisenberg model when $\Delta = 1$ and to the Ising model in the extreme anisotropy limit, when $\Delta = 0$. 

      The \gls{TM} ions of our systems lie on an almost perfectly planar octagonal ring. Given their annular shape, it seems appropriate to express the Hamiltonian in cylindrical coordinates, with the axial direction $z$ taken perpendicular to the molecular plane, and the in-plane spin components decomposed into tangential and radial terms, 
      $\mathbf{\hat{S}}_i = (\hat{S}^t_i,\hat{S}^r_i,\hat{S}^z_i)$. 
      Within this cylindrical representation of the localized spins it also seems reasonable to assume that most of the exchange interactions are captured by a diagonal Hamiltonian: 

      \begin{align}
          \hat H_\mathrm{ex} 
            = & \, 
            \sum_i 
            J^t \hat S_i^t\hat S_{i+1}^t + J^r \hat S_i^r\hat S_{i+1}^r + J^z \hat S_i^z\hat S_{i+1}^z 
            \label{cex1} \\
              = & \,
              \sum_{\alpha=t,r,z}
              \sum_i  
              J^{\alpha} \hat{S}_i^{\alpha} \hat{S}_{i+1}^{\alpha}
          \notag 
      \end{align}

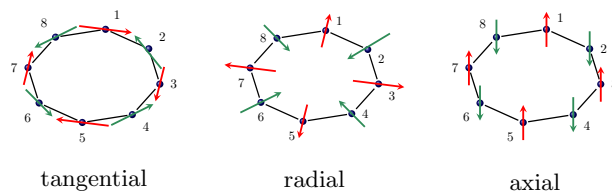
\begin{figure}[!b]
          \centering
          \begin{tikzpicture}
         
              \node (img1) {
                \includegraphics[width=2.5cm]{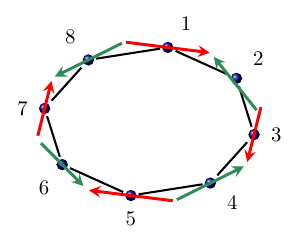}
              };
              \node[below=0mm of img1.south] {tangential};
         
              \node[right=0.5mm of img1] (img2) {
                \includegraphics[width=2.8cm]{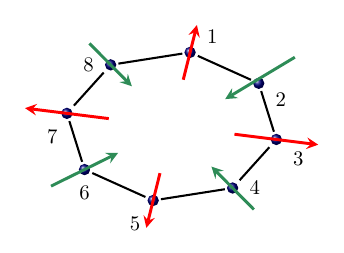}
              };
              \node[below=0mm of img2.south] {radial};
      
              \node[right=0.5mm of img2] (img3) {
                \includegraphics[width=2.5cm]{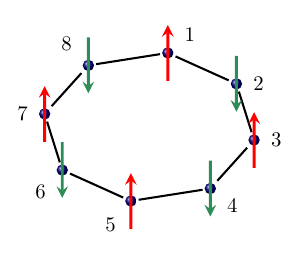}
              };
              \node[below=0mm of img3.south] {axial};
          
          \end{tikzpicture}
          \caption{The three spin \gls{AFM} non-collinear reference configurations.}
          \label{fig:non-collinear-config}
      \end{figure}

      Obviously, this expression is not compatible with the diagonal form of the exchange Hamiltonian in Cartesian coordinates, Eq. \eqref{heis2}, and a form-invariant expression would require the inclusion of off-diagonal terms, at least between the $x$ and $y$ components. The type and relevance of these terms will be discussed later, when presenting other kinds of couplings between spins. For the moment we will assume a diagonal expression for the exchange interactions between neighbor spins and will compute the corresponding couplings one by one. A first estimate of $J^t$, $J^r$, and $J^z$ can be obtained by directly comparing the energy of the \gls{AFM} configurations in the three directions, shown in Fig. \ref{fig:non-collinear-config}, with that of the corresponding \gls{FM} orderings. The couplings obtained within this approach are shown in Table \ref{tab:ncoll-ring}.

      \begin{table}[!h]
      \centering
            \begin{tabular}{ l|r|r|r|r } 
            \hline
          $J_1$(meV) &\multicolumn{2}{c|}{ \ce{Cr8}-ring}
            &\multicolumn{2}{c}{ \ce{V8}-ring}\\
            \hline
            Parameter	&+U &	+U+V	&+U &	+U+V\\
            \hline
      $ J^r $&	0.596& 0.848 & -0.643 & -0.403 \\
      $ J^t $&	0.596& 0.848 & -0.654 & -0.403 \\
      $ J^z $&	0.843& 1.198 & -0.913 & -0.585 \\ \hline
      $ J^\mathrm{coll} $&	0.843& 1.199 & -0.917 & -0.577 \\
        \hline  
      \end{tabular}
      \caption{Exchange couplings from non-collinear \gls{DFT} calculations using \gls{LDA}+U and \gls{LDA}+U+V functionals. For comparison, the collinear results $ J^\mathrm{coll} $ from Ref. \cite{stocco2025} are also included.} 
          \label{tab:ncoll-ring}
      \end{table}

The axial $z$ components show excellent agreement with the collinear values reported in Ref. \cite{stocco2025} for both structures. 
In-plane components are somewhat smaller for both  systems, although still of the same order of magnitude. It is worth noting that, consistently with the systems' geometry, these nearest-neighbor tangential and radial components have approximately the same values; in addition, the couplings become higher (i.e. less negative for \ce{V8}) along all directions when the extended \gls{DFT}+U+V scheme is employed. 

\begin{figure}[!t]
 \centering
\includegraphics[width=1.0\linewidth]{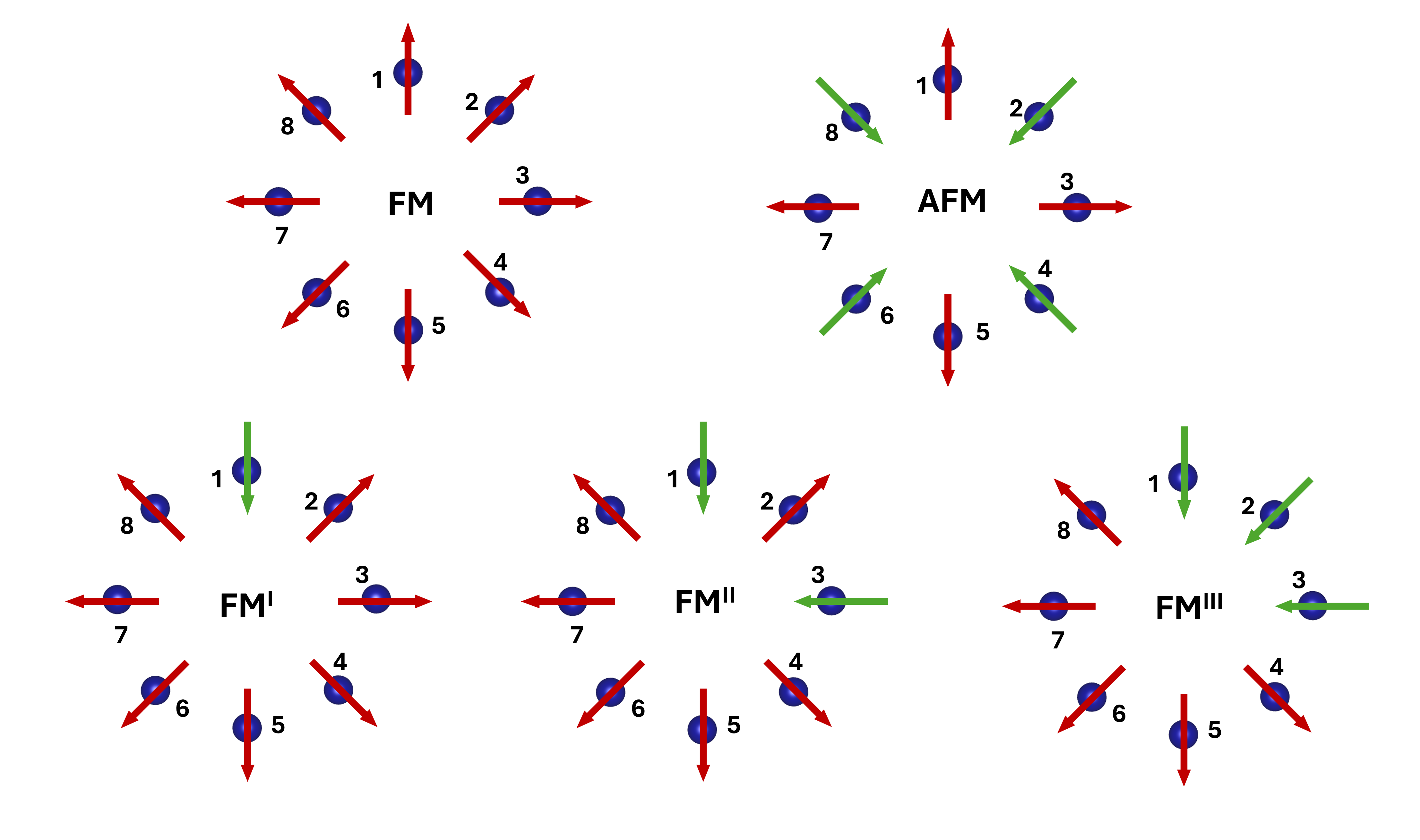}
\caption{Non-collinear magnetic setups, in V$_8$ radial configuration, considered to calculate the exchange parameters between all neighbor magnetic sites, $J_2$, $J_3$ and $ J_4$ .}
  \label{fig:fig4}
\end{figure} 

Consistently with Ref. \cite{stocco2025}, to evaluate the couplings between beyond–nearest neighbors (named, respectively, $J_{2}$, $J_{3}$, and $J_{4}$), we considered three additional spin configurations, named \gls{FM}$^i$, (where $ i=I,II,III $ indicates the number of spins reversed relative to the \gls{FM} configuration), which are sketched in Fig. \ref{fig:fig4} for the radial orientation. According to the procedure consolidated in Ref. \cite{stocco2025} and briefly refreshed at the end of Section \ref{sec:03}, by mapping the \gls{DFT} total energies on the predictions of an extended Heisenberg model with exchange interactions up to fourth neighbors, we obtain the following equations, valid in all three directions:  

      \begin{equation}
      \begin{aligned}
      &E_{\rm FM-AFM}		=2S^{2}(8J_{1}+8J_{3})&\\
      &E_{\rm FM-FM^{I}}	=2S^{2}(2J_{1}+2J_{2}+2J_{3}+J_{4}) &\\
      &E_{\rm FM-FM^{II}}	=2S^{2}(4J_{1}+2J_{2}+4J_{3}+2J_{4})&\\
      &E_{\rm FM-FM^{III}}	=2S^{2}(2J_{1}+4J_{2}+6J_{3}+3J_{4}). 
      \end{aligned}
      \end{equation}

      By solving this system of  equations we extract the exchange couplings in the axial, radial and tangential directions up to the fourth neighbor (the furthest possible on an eight-atom ring). The values obtained from non-collinear \gls{DFT}+U and \gls{DFT}+U+V calculations are shown in Table \ref{table:V8-J-fit-non-collinear} and Table \ref{table:V8-J-fit-non-collinear+V}, respectively.

\begin{table}[!htb]
   \centering
  \begin{tabular}{l|r|r|r} 
        \hline
     LDA+U &\multicolumn{3}{c}{ \ce{Cr8}-ring }\\
        \hline
     Parameter	& $J_2$ (meV) &$J_3$ (meV)& $J_4$ (meV)	\\
        \hline
      $ J^r $& -0.0003 &  $\ll J_2$ &  $\ll J_2$\\
      $ J^t $& -0.0005 &  $\ll J_2$ & $\ll J_2$ \\
      $ J^z $& -0.0150 &  $\ll J_2$ &  $\ll J_2$\\
        \hline
     LDA+U &\multicolumn{3}{c}{ \ce{V8}-ring }\\
        \hline
     Parameter	&$J_2$ (meV) &$J_3$ (meV)& $J_4$ (meV) \\
        \hline
      $ J^r $&0.120 &-0.020 & 0.024\\
      $ J^t $&0.163 &-0.017 & -0.019\\
      $ J^z $&0.207 & 0.013 & 0.001\\
        \hline
   \end{tabular}
\caption{$J_{2}$, $J_{3}$ and $J_{4}$ exchange couplings from non-collinear \gls{DFT}+U (\gls{LDA}+U) calculations.}
\label{table:V8-J-fit-non-collinear}
\end{table}

\begin{table}[!htb]
   \centering
     \begin{tabular}{ l|r|r|r} 
    \hline
     LDA+U+V &\multicolumn{3}{c}{ \ce{Cr8}-ring }\\
    \hline
     Parameter	& $J_2$ (meV) &$J_3$ (meV)& $J_4$ (meV)	\\
    \hline
      $ J^r $& 0.0038  & 0.0057 &  0.0004\\
      $ J^t $& 0.0045  & 0.0058 & -0.0004 \\
      $ J^z $& -0.0016 & 0.0002 & 0.002\\
        \hline
     LDA+U+V &\multicolumn{3}{c}{ \ce{V8}-ring }\\
            \hline
     Parameter	&$J_2$ (meV) &$J_3$ (meV)& $J_4$ (meV) \\
            \hline
      $ J^r $&0.178 &-0.024 & -0.028\\
      $ J^t $&0.177 &-0.022 & -0.020\\
      $ J^z $&0.262 &-0.037 & 0.066\\
        \hline
      \end{tabular}
\caption{$J_{2}$, $J_{3}$ and $J_{4}$ exchange couplings from non-collinear \gls{DFT}+U+V (\gls{LDA}+U+V) calculations. } 
\label{table:V8-J-fit-non-collinear+V}
 \end{table}

From a rapid inspection of the effective longer-range couplings shown in Tables \ref{table:V8-J-fit-non-collinear} and \ref{table:V8-J-fit-non-collinear+V} it is easy to realize some important facts.
For \ce{Cr8} only $J_2^z$ from \gls{LDA}+U is appreciable (of the order of 1-2 \% of $J^z_1$) while all the other couplings turn out to be at least one order of magnitude smaller (with occasional change of sign), and thus completely negligible. For \ce{V8}, instead, $J_2$ are a significant fraction of $J_1$, both from \gls{LDA}+U and \gls{LDA}+U+V, while $J_3$ and $J_4$ are about one order of magnitude smaller than $J_2$. It is also interesting to note that $J_2$ are all positive, thus revealing a tendency to establish an \gls{AFM} order between second nearest neighbor spins that competes with the dominant \gls{FM} one between first neighbors, imposed by $J_1$. This trend was already found from our collinear-spin calculations, presented in Ref. \cite{stocco2025}. 

A comparison between the two sets of calculations, especially for the $J_2$ couplings of \ce{V8}, reveals that \gls{LDA}+U+V generally yields higher values (in modulus) for  the effective interactions. Furthermore, while differentiating the $z$ components from the planar ones, it brings the values of the $r$ and $t$ components closer to each other, thus re-establishing a Hamiltonian that reflects more closely the cylindrical symmetry of these annular systems.

\subsection{Biquadratic terms} \label{bqt}

A second prerogative enabled by non-collinear spin calculations concerns the possibility to \textit{rotate} single spins from a  reference configuration (e.g., \gls{AFM} or \gls{FM}) of the system. The corresponding energetics allows not only to verify the models in Eqs. \eqref{heis2} and \eqref{cex1} (e.g., ascertain that off-diagonal couplings are negligible) but also to reveal other aspects of the magnetic behavior of the systems as, for example, possible anisotropies and other kinds of interactions.

\begin{figure*}[!htb]
 \centering
  \includegraphics[width=1.0\textwidth]{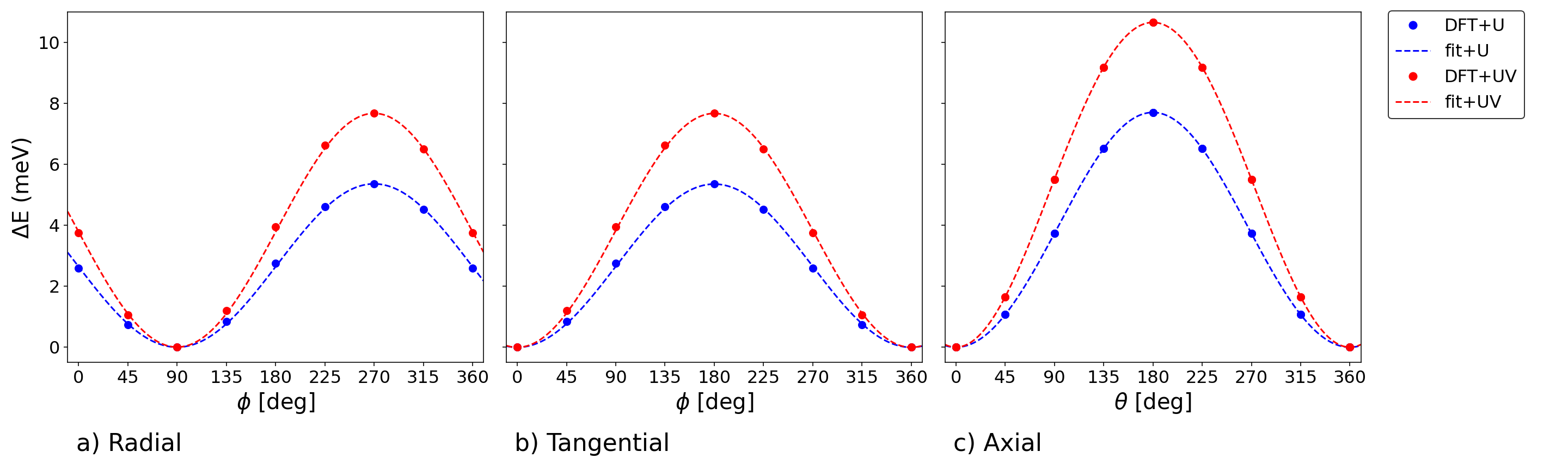}
  \caption{Non-collinear \gls{DFT} total energy profiles obtained rotating a single spin in Cr$_8$ starting from the \gls{AFM} ground state configuration in each direction. Dots are results from ab initio calculations; dashed curves are fit obtained from Eq. (\ref{BQ3}).}
 \label{fig:non-collinear-Cr8} 
    \end{figure*}
          
\begin{figure*}[!htb]
  \includegraphics[width=1.0\textwidth]{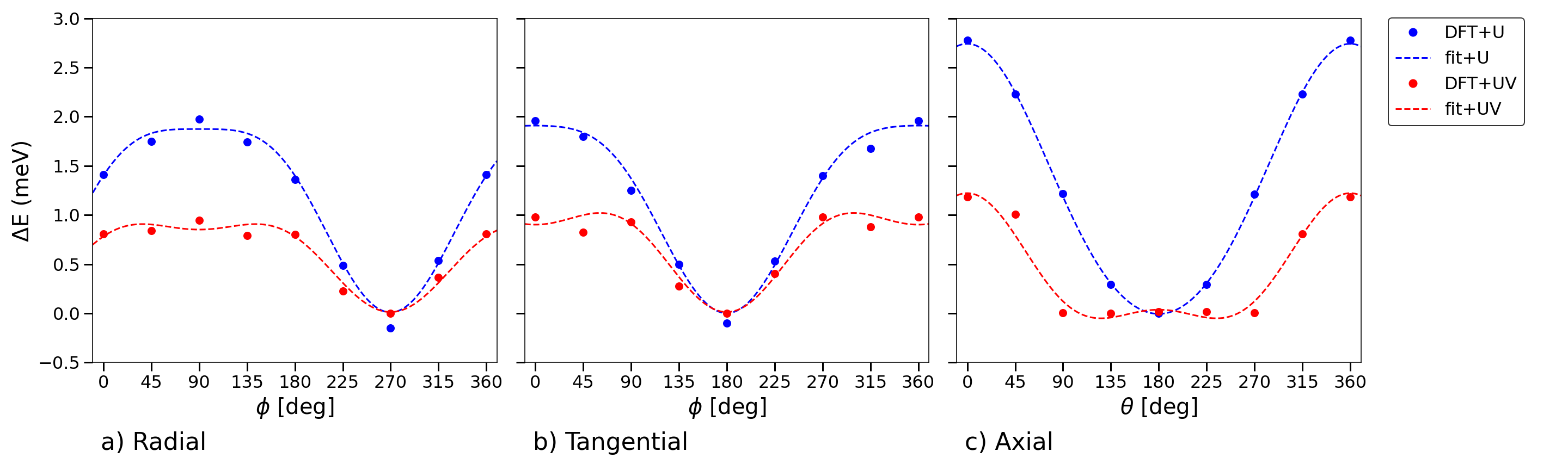}
  \caption{Non-collinear \gls{DFT} total energy profiles obtained rotating a single spin in V$_8$ starting from the \gls{FM} ground state configuration in each direction. Dots are results from ab initio calculations; dashed curves are fit obtained from Eq. (\ref{BQ3}).}
   \label{fig:non-collinear-V8} 
   \end{figure*}
          
The results of calculations where a single spin is rotated are in Figs. \ref{fig:non-collinear-Cr8} and \ref{fig:non-collinear-V8} for \ce{Cr8} and \ce{V8}, respectively. The graphs report the total energy change as a function of the spin rotation angle with respect to  reference configurations with all spins aligned, respectively, along the radial, tangential and axial directions. The angles $\theta$ (polar) and $\phi$ (azimuthal) are measured from the $z$ axis along a $rz$ plane, and from the tangential direction in the $xy$ plane, respectively. The rotation from the axial direction ($\theta = 0$) along a $tz$ is almost exactly equivalent with the one on $rz$ and is not shown. 
All graphs are zeroed at their minimum and compare the results of direct \gls{DFT} calculations (dots) with the prediction of a fit obtained from Eqs. (\ref{BQ3}) (dotted curves, see discussion below).  
Some of the \ce{V8} curves deviate significantly from a sinusoidal form, revealing that the spin Hamiltonian parametrized as in Eq. (\ref{Hex1}) is insufficient to describe the magnetic energy landscape and necessitates the inclusion of additional interaction terms.

Deviations from a sinusoidal behavior are often attributed to on-site anisotropies, which arise from the interaction of localized spins with the local crystal field. However, these couplings are normally a consequence of \gls{soc}, which is not included in our calculations. Therefore, in order to explicitly rule out this possibility -- i.e. the occurrence of spurious \gls{soc}-like artifacts as a consequence of approximate \gls{xc} functionals --, we performed a separate calculation in which seven V atoms out of eight were substituted by Sc, an element which is non-magnetic in its 3+ state. 
By rotating the spin of the remaining V center, we obtained an almost completely flat energy profile, as illustrated in Appendix \ref{test}. This outcome confirms the absence of spurious on-site anisotropies, and points towards higher order spin-spin interactions to explain the peculiar shape of the energy profiles in Fig. \ref{fig:non-collinear-V8}.

The additional terms considered here to complement the exchange interactions are the so-called \textit{\gls{BQ} couplings}, that introduce a dependence on the {\it square} of the scalar product of spins on neighbor sites. After their introduction, the  spin Hamiltonian reads: 
     
      \begin{eqnarray}
        \hat H &\equiv& \hat H_\mathrm{ex} + \hat H_\mathrm{BQ} \notag \\
        &=& 
          \sum_i 
          \sum_{\alpha=t,r,z}
\left[ J^{\alpha}
\hat{S}_i^{\alpha} \hat{S}_{i+1}^{\alpha} +
J_\mathrm{BQ}^{\alpha}
( \hat{S}_i^{\alpha} \hat{S}_{i+1}^{\alpha} )^2 \right]
        \label{hspinbq}
      \end{eqnarray}

where, for the sake of simplicity, a diagonal expression has been adopted. 
The \gls{BQ} terms have already been discussed in the literature, e.g. as possible sources of a non-Heisenberg-like behavior in some oxides as, for example, perovskite manganites \cite{fedorova15}. In general, they are expected to be smaller than Heisenberg bilinear exchange interactions, as they can be understood as higher-order terms in a perturbative expansion around the ground state configuration. However, they can significantly influence  quantities such as magnetic anisotropies \cite{xgao16}, spin-wave gaps and topological spin excitations \cite{biqu,biqu2}.
      
      The \glspl{BQ} have been extracted from fitting our non-collinear spin \gls{DFT} total energies with the predictions of a Hamiltonian that generalizes the one in Eq. \eqref{hspinbq} by letting exchange and \gls{BQ} couplings be different in the three cylindrical directions. By rotating single spins from  reference configurations with all the momenta aligned along the $r$, $t$,  and $z$ directions, we obtain the following expressions for the energy as a function of the rotation angles: 

      \begin{equation}
      \begin{split}
        E^r(\phi)   &= -2 S^2 (J_1+J_2)^r\sin{\phi} + 2 S^4 J_\mathrm{BQ}^r \sin^2{\phi}\\
        E^t(\phi)   &= -2 S^2 (J_1+J_2)^t\cos{\phi} + 2 S^4 J_\mathrm{BQ}^t \cos^2{\phi} \\
        E^z(\theta) &= -2 S^2 (J_1+J_2)^z\cos{\theta} + 2 S^4 J_\mathrm{BQ}^z \cos^2{\theta}.  
      \end{split}
      \label{BQ3}
      \end{equation}

These equations are used to fit the ab initio results in Figs. \ref{fig:non-collinear-Cr8} and \ref{fig:non-collinear-V8}, where a comparison is made with the results of single \gls{DFT}+U or \gls{DFT}+U+V calculations.
In extracting the \glspl{BQ}, the exchange terms are held fixed to the values shown in Tables \ref{tab:ncoll-ring}, \ref{table:V8-J-fit-non-collinear} and \ref{table:V8-J-fit-non-collinear+V}, while off-diagonal terms are assumed to be negligible also in the \gls{BQ} sector.  
From a comparison between Figs. \ref{fig:non-collinear-Cr8} and \ref{fig:non-collinear-V8}, it may seem counterintuitive that the energy of the system exhibits wider oscillations when computed with \gls{DFT}+U+V than \gls{DFT}+U for \ce{Cr8}, while the opposite is true for \ce{V8}. However, it should be kept in mind that these energies contain relatively large contributions from exchange ($J_1$ and $J_2$ - see Eqs. \ref{BQ3}), quite different for the two systems, and that the reference configurations are also different (\gls{FM} for \ce{V8}, \gls{AFM} for \ce{Cr8}); these facts easily justify the observed differences.

The \glspl{BQ} couplings obtained from fitting the ab initio energies with the expressions in Eqs. \ref{BQ3} are shown in Table \ref{tab:biqu}.

\begin{table}[!b]
\centering
\begin{tabular}{ c|c|c|c|c } 
            \hline
   &  \multicolumn{2}{c|}{LDA+U} & \multicolumn{2}{c}{LDA+U+V} \\
           \hline
  Parameter	&\ce{Cr8}-ring & \ce{V8}-ring & \ce{Cr8}-ring & \ce{V8}-ring\\
      \hline
      $ J_{\mathrm{BQ}}^r $&	0.004&	-0.230 & 0.005 & -0.175\\
      $ J_{\mathrm{BQ}}^t $&	0.002&	-0.208 & 0.0002 & -0.227\\
      $ J_{\mathrm{BQ}}^z $&	0.012&	 0.091 & -0.018 & 0.260\\
      \hline
      \end{tabular}
 \caption{\gls{BQ} terms (in meV) obtained from the fit of non-collinear spin \gls{LDA}+U and \gls{LDA}+U+V calculations.}
     \label{tab:biqu}
      \end{table}

Consistently with the qualitative behavior observed in Figs. \ref{fig:non-collinear-Cr8} and \ref{fig:non-collinear-V8}, while \ce{Cr8} exhibits almost vanishing \gls{BQ} couplings (reaching at most 0.012 meV and -0.018 meV along the axial direction for \gls{LDA}+U and \gls{LDA}+U+V, respectively) \ce{V8} is characterized by stronger interactions that for \gls{LDA}+U (\gls{LDA}+U+V) reach -0.230 (-0.175) and -0.208 (-0.227) meV in the radial and tangential directions, respectively, and 0.091 (0.260) meV along the $z$ axis. 

These results indicate that the \glspl{BQ} in \ce{V8} favor spin alignment on the molecular plane, while disfavor it along the perpendicular direction, thus contrasting the trend emerging from the exchange interactions. In addition, the difference between radial and tangential components introduces an element of in-plane anisotropy that is not present in the exchange sector of the Hamiltonian, $\hat H_\mathrm{ex}$. These trends are similar between \gls{LDA}+U and \gls{LDA}+U+V, although \gls{LDA}+U+V promotes higher values of the \gls{BQ} couplings along z, that become prevalent (in modulus) on the other components.    
      
In concluding this Section it is appropriate to remark that the introduction of the \glspl{BQ} in the Hamiltonian does not invalidate the calculation of the exchange couplings described in the previous section. 
In fact, the latter were obtained starting from the spin configurations shown in Fig. \ref{fig:fig4}, by inverting the direction of single spins. In view of Eq. \eqref{BQ3}, this corresponds to a polar angle rotation of $\theta=180^\circ$, for which the squared cosine factor suppresses the \gls{BQ} couplings $J^z_\mathrm{BQ}$ in the energy differences.

\subsection{Dzyaloshinskii-Moriya interactions}\label{sec:DM}
Non-collinear spin \gls{DFT} calculations open the possibility to consider also more exotic configurations of  localized moments and to evaluate the entity of the magnetic interactions that tend to stabilize them. An example is provided by \gls{DM} interactions that are antisymmetric  with respect to the exchange of the involved spins and tend to stabilize helical orders (e.g., spin spirals). A practical way to represent these interactions is through the vector product of neighbor spins:
      \begin{equation}
          \hat H_\mathrm{DM} = \sum_i {\mathbf d} \cdot \left(\hat{\mathbf S}_i \times \hat{\mathbf S}_{i+1}\right).
          \label{dm1}
      \end{equation}

      The vector ${\mathbf d}$ characterizes the strength and orientation of the \gls{DM} interactions. 
      In cylindrical coordinates, ${\mathbf d}$ = $(d^r,d^t,d^z)$, and  Eq. \eqref{dm1} can be equivalently rewritten as:

      \begin{equation}
      \hat H_\mathrm{DM} =  
          \sum_i \hat{\mathbf S}_i \cdot \textbf{J}_\mathrm{DM} \cdot \hat{\mathbf S}_{i+1}
      \label{dmj}
      \end{equation}
      where the antisymmetric exchange coupling $J_\mathrm{DM}$ can be represented through the following antisymmetric matrix:
      \begin{align}
          \textbf{J}_\mathrm{DM} =  \,
          \begin{bmatrix}
              0 & d^z   & -d^t \\
          -d^z  & 0     & d^r  \\
          d^t  & -d^r  & 0  \\
          \end{bmatrix}.
          \label{dm2}
      \end{align}

      The $\textbf{J}_\mathrm{DM}$ expression in Cartesian coordinates can be easily obtained by substituting $r$ and $t$ components with $x$ and $y$.

Due to its antisymmetric character, for $\hat H_\mathrm{DM}$ to give a non-zero contribution to the energy, it is necessary that the angle between the $i^{\text{th}}$ and $\left(i-1\right)^{\text{th}}$ spins is not the opposite of the one between the $\left(i+1\right)^{\text{th}}$ and $i^{\text{th}}$, while both need to be different from 0$^{\circ}$ and 180$^{\circ}$. This observation helps clarify how \gls{DM} interactions can stabilize helical spin configurations. It also elucidates that \gls{DM} couplings provide vanishing contributions to the energy when single spins are rotated from a reference collinear configuration (thus forming opposite angles with the spins on either side) or when the spins on neighbor sites are parallel or antiparallel (thus nullifying the vector product). Therefore, the presence of \gls{DM} interactions does not compromise the calculation of exchange or \gls{BQ} couplings as it was performed in this work, nor could be revealed by any of them. 

In general, the origin of \gls{DM} couplings between localized spins is traced back to \gls{soc} \cite{DZYALOSHINSKY1958241, Moriya}. However, based on the absence of heavy elements in the considered systems, all the calculations presented in this work were performed without \gls{soc}. 
Nevertheless, by rewriting a diagonal exchange Hamiltonian from Cartesian to cylindrical coordinates, it is quite straightforward to achieve, for a ring-shaped system, finite interaction terms of the \gls{DM} type (e.g., consistent with  Eq. \eqref{dm1}). This can be  shown by focusing on the in-plane ($xy$) part of the diagonal exchange Hamiltonian in Eq. \eqref{heis2} for the $i^{th}$ site: 

      \begin{widetext}
          \begin{eqnarray}
          \label{planeJ}
          &&\sum_{\alpha = x,y} \frac{J}{2} \left(S_i^{\alpha} S_{i+1}^{\alpha} + S_i^{\alpha} S_{i-1}^{\alpha}\right) = \nonumber \\
          &&\frac{J}{2}  \left[\left(S_i^r S_{i+1}^r + S_i^r S_{i-1}^r\right) + 
          \left(S_i^t S_{i+1}^t + S_i^t S_{i-1}^t\right) \right]\cos{\theta} + 
          \frac{J}{2}  
          \left[\left(-S_i^r S_{i+1}^t + S_i^r S_{i-1}^t\right) + 
          \left(S_i^t S_{i+1}^r - S_i^t S_{i-1}^r\right) \right] \sin{\theta} 
          = \nonumber \\
          &&\frac{J}{2} \cos{\theta} \left[\left(S_i^r S_{i+1}^r + S_i^r S_{i-1}^r\right) + 
          \left(S_i^t S_{i+1}^t + S_i^t S_{i-1}^t\right) \right] + 
          \frac{J}{2} \sin{\theta} 
          \left[\left(S_i^t S_{i+1}^r - S_i^r S_{i+1}^t\right) + 
          \left(S_{i-1}^t S_i^r  - S_{i-1}^r S_i^t \right) \right].
          \end{eqnarray}
      \end{widetext}

where $\theta$ is the angle between adjacent radial directions and $\theta=45^{\circ}$ for the molecules of octahedral shape considered in this work. The last line of Eq. \eqref{planeJ} is quite self-explanatory: while the first term contains diagonal  exchange interactions ($r$ and $t$), the second is  antisymmetric with respect to the exchange of site $i$ and $i+1$ or $i-1$ and can be assimilated to the $z$ component of Eq. \eqref{dm1} with $d^z = \frac{J}{2} \sin{\theta}$. A similar (reciprocal) result is obtained starting from a nondiagonal Cartesian \gls{DM} term and rewriting it in cylindrical coordinates. While this seems to break the covariant nature of the Hamiltonian, a more general $\hat H$ containing both exchange and \gls{DM} interactions terms will transform under a change of basis into an operator containing again exchange and \gls{DM} interactions terms. 
      This observation suggests that a minimal covariant  formulation of the spin Hamiltonian should contain interactions of \textit{both} exchange and antisymmetric \gls{DM} type.

The \gls{DM} interactions were historically introduced as a relativistic correction to the exchange Hamiltonian $\hat H_\mathrm{ex}$  and are generally expected to be present only when \gls{soc} effects are substantial. 
Eq. \eqref{planeJ} suggests that finite \gls{DM} interactions can arise from rewriting $\hat H_\mathrm{ex}$ in cylindrical coordinates, in the presence of geometrical curvature. 
This result seems consistent with existing literature reporting on the emergence of Rashba-type \gls{soc} in systems with a curved shape \cite{gentile13,gentile15}
      or as a consequence of curvature-induced charge- or spin-currents \cite{spin-curr,Sadamichi,PhysRevLett.116.247201,PhysRevLett.121.147203}, even in ground states where atomic \gls{soc} effects are negligible \cite{cardias2020dzyaloshinskii,Cardias2020}. 
Quite interestingly, however, our calculations are based on \gls{xc} functionals that contain no explicit account of currents of any type.

In order to understand whether the \gls{DM} interactions found here result exclusively from a change of basis or rather are, to some extent, intrinsic to the considered systems, we proceed here to their direct calculation. 
For a precise assessment of the role of their curvature, we compare the \gls{DM} interactions of the rings with those of the corresponding linear chains, shown in Fig. \ref{fig:fig2-a}. This comparison was already proposed in other works aimed at assessing the dependence of effective exchange interactions on bond angles \cite{Bellini2008}. 

\begin{figure}[!t]
        \centering
   \includegraphics[width=0.9\linewidth]{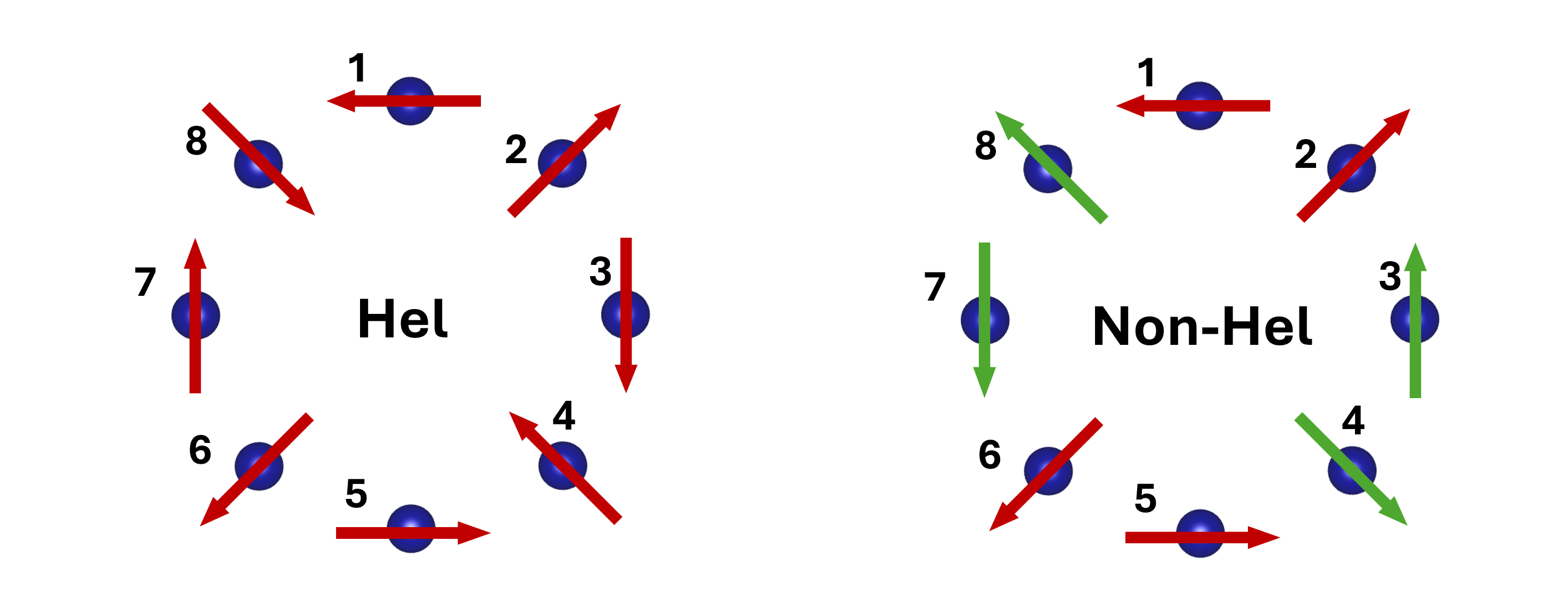}
\caption{Magnetic spin configurations used to evaluate d$^z$ in \ce{Cr8} and \ce{V8}}
          \label{fig:dm-config}
      \end{figure}

For the \gls{DM} calculations we  compare two families of non-collinear spin arrangements: helical and non-helical. As an illustrative example, Fig. \ref{fig:dm-config} shows the spin configurations used to determine $d^z$ (i.e., the coupling between the in-plane components of the spins). Similar arrangements were employed also to evaluate $d^r$ and $d^t$.  
Based on the expression of the exchange and \gls{DM} interactions, the effective couplings were obtained from the \gls{DFT} total energy differences between helical and non-helical configurations using the following formulas:
      \begin{equation}
      \begin{split}
      \Delta E^{rt} = - 8  S^2 \cdot ( J_2^r + J_2^t )  - 8 S^2 \cdot d^z \\
      \Delta E^{zt} = - 8  S^2 \cdot ( J_2^z + J_2^t )  - 8 S^2 \cdot d^r \\
      \Delta E^{zr} = - 8  S^2 \cdot ( J_2^z + J_2^r )  - 8 S^2 \cdot d^t. 
          \label{DM}
          \end{split} 
      \end{equation}

\begin{table}[!t]
\centering
      \begin{tabular}{ l|r|r|r|r } 
        \hline
   &  \multicolumn{2}{c|}{\ce{Cr8}-ring } & \multicolumn{2}{c}{\ce{V8}-ring }\\
        \hline
 Parameter &+U  &+U+V &+U  &+U+V   \\
        \hline
    $d^r$  & -0.193 & -0.007 &  -0.381 & -0.433 \\
    $d^t$  & -0.193 & -0.007 &  -0.338 & -0.433 \\
    $d^z$  &  0.594 & 0.853  &  -0.515 & -0.939 \\
        \hline
    \end{tabular}
\caption{\gls{DM} interaction parameters of \ce{Cr8} and \ce{V8}-rings  with \gls{LDA} + U and  \gls{LDA} + V. All values are in meV.}
\label{Tab:dm-rings} 
      \end{table}

As can be easily realized, only second-nearest neighbor exchange interactions ($J_2$) enter the above expressions; nearest-neighbor coupling contributions  do not provide any contribution due to the orientation of spins along orthogonal directions on neighbor sites. 

      The results of our calculations are reported in Table \ref{Tab:dm-rings}, which makes a comparison between \gls{LDA}+U and \gls{LDA}+U+V functionals for both systems.

      Table \ref{Tab:dm-rings} shows that the dominant \gls{DM} interactions are, for both systems, along the $z$ axis and grow in modulus with the addition of inter-site Hubbard interactions (i.e., from +U to +U+V). The main difference between \ce{Cr8} and \ce{V8} is in the sign of these couplings which is positive for the former and negative for the latter (thus lowering the energy of helical spin orders).
      The numerical prevalence of $d^z$ over the in-plane components is consistent with the adoption of cylindrical coordinates where in-plane $x$ and $y$ components need to be mixed (with transformations keeping $z$ fixed) in order to achieve $r$ and $t$ (see Eq. \eqref{planeJ}). On the other hand $d^r$ and $d^t$ are in general non-zero (except for \ce{Cr8} with U+V) and negative, thus favoring the misalignment of spins from the $z$ direction.  These in-plane couplings are particularly consistent for \ce{V8} suggesting, for this system, an intrinsic (i.e., non exclusively geometrical) origin. 

      In order to complete the above discussion and further detail the distinction between geometrical and intrinsic \gls{DM} interactions, we now turn  to the calculation of \gls{DM} couplings for linear chains. 
      Specifically, we study linear molecules of four magnetic sites, as this is the minimal length to construct helical spin configurations. 

\begin{figure}[!t]
     \centering
   \includegraphics[width=\linewidth]{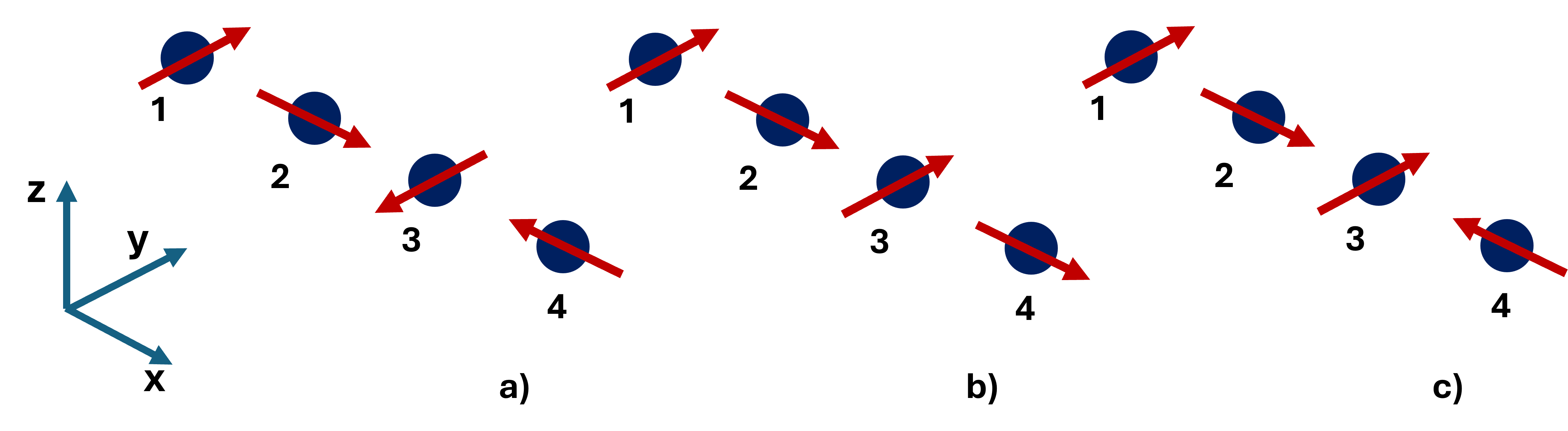}
\caption{Helical and non-helical spin configurations used to evaluate the $d^z$ antisymmetric exchange couplings in the chain model. }
\label{fig:d_config_chain}
      \end{figure}

      Due to the linear geometry of these systems,  we adopt Cartesian coordinates that distinguish a longitudinal direction -- along the chain, denoted $x$ -- from the transverse one. Fig. \ref{fig:d_config_chain} compares the three spin configurations used to evaluate the \gls{DM} interactions (note that ``a'' has helical character while ``b'' and ``c'' do not). The configurations have spins in the $xy$ plane and are specifically used to calculate $d^z$; to evaluate $d^x$ and $d^y$ analogous arrangements are used, but with spins on the $yz$ and $xz$ planes, respectively. 

      The \gls{DM} parameters along the $i^{th}$ direction can be extracted from the following relation involving the total energies of the  spin configurations, labeled  as in Fig. \ref{fig:d_config_chain}:

      \begin{equation}
      E_a = (2 \cdot E_c - E_b )  - 4 S^2 \cdot d^i.
          \label{DM-chain}
          \end{equation}

      The results for the $d^i$ in the \ce{Cr4} and \ce{V4} chains are shown in Table \ref{Tab:dm-chains}.

\begin{table}[!t]
\centering
      \begin{tabular}{ l|r|r|r|r } 
      \hline
  &    \multicolumn{2}{c|}{\ce{Cr4}-chain } &  \multicolumn{2}{c}{\ce{V4}-chain } \\
       \hline
    Parameter      & +U     &+U+V & +U     &+U+V    \\
    \hline
    $d^x$& 1$\times 10^{-4}$   &-0.010 & -0.157   & -0.077 \\
    $d^y$& 1$\times 10^{-4}$   &-0.010 & -0.041   & -0.155 \\
    $d^z$& -1$\times 10^{-4}$  &-0.010 & 0.400    & -0.055 \\
  \hline
  \end{tabular}
\caption{\ce{Cr4}-chain and \ce{V4}-chain \gls{DM} interaction parameters with \gls{LDA}+U and \gls{LDA}+U+V. All values are in meV. }
\label{Tab:dm-chains}
      \end{table}

Importantly, all the components of the \gls{DM} couplings of the \ce{Cr4} chain are negligible, with only minor effects deriving from the introduction of inter-site Hubbard couplings. The \ce{V4} chain presents, instead, larger interactions that are in most cases of the order of 10$^{-2}$ - 10$^{-1}$ meV. Within \gls{LDA}+U the largest \gls{DM} coupling is $d^z$ that reaches 0.4 meV. \gls{DFT}+U+V predicts, instead, $d^y$ as the largest coupling (-0.155 meV) with a general decrease of all the (absolute) values and a change in sign of $d^z$ that goes from 0.4 to -0.055 meV. Quite interestingly, neither set of \gls{DM} interactions seems to be consistent with the geometry of the system (e.g., similar components along $y$ and $z$ directions, transverse to the axis of the chain), possibly reflecting an enhanced sensitivity of the \gls{DM} interactions to the specific orientation of the \gls{TM}-centered octahedra around the longitudinal axis.
      Comparing these results with those in Table \ref{Tab:dm-rings}, it is  clear that for the Cr-based systems the \gls{DM} interactions are to be traced back to the structural curvature imposed by its ring shape and to the rewriting of the exchange interactions in cylindrical coordinates, as shown in Eq. \eqref{planeJ}. This observation, corroborated also by the vanishingly small values of \gls{DM} interactions obtained in the chain, is particularly true within \gls{DFT}+U+V that leads to non-zero \gls{DM} couplings only along the direction ($z$) perpendicular to the plane of the \ce{Cr8} ring.
  
      The case of V-based systems is, instead, quite different. The non vanishing values of \gls{DM} interactions obtained for the \ce{V4} chain seem to confirm that these interactions cannot be entirely explained by geometrical arguments, nor arise from a rewriting of exchange in cylindrical coordinates. On the contrary, they support an intrinsic origin of \gls{DM} interactions (probably related to the electronic structure of the V$^{3+}$ centers) that leads to numerically consistent $d^r$ and $d^t$ in the \ce{V8} ring (while $d^z$ is still prevalent) even without significant mixing of spin coordinates out of the molecular plane. However, the scarce adherence of the \gls{DM} interactions with the geometry of the \ce{V4} chain and the sensitivity of their values to the approximation used are still puzzling aspects. While their non-zero value seems a robust result, whether the delicate convergence (in the self-consistent solution of Kohn-Sham equations) shown in this case by non-collinear spin calculations can be at least part of the explanation remains to be understood. 
\\
      In summary, our results for the V-based systems demonstrate the possibility of finite \gls{DM} interactions even without atomic \gls{soc} (in fact absent in our calculations). However, the microscopic (electronic) mechanisms responsible for their occurrence remain quite unclear. 
      
Interesting hypotheses to be explored include: \textit{i}) the idea that finite \gls{DM} interactions could also be the effect of the less-than-half-filled $d$ subshells of the V magnetic centers, as demonstrated in our previous work for the anomalous \gls{FM} character of the exchange couplings \cite{stocco2025}; 
\textit{ii}) the relationship of \gls{DM} interactions with charge and spin currents, possibly appearing also in the ground state and responsible of significant spin-orbit torques \cite{meyer23,spin-curr,Sadamichi,Cardias2020,PhysRevLett.116.247201}. 
The evaluation of these possibilities and of their possible interrelations is left for future work.

\subsection{
Spin Hamiltonian diagonalization: magnetic susceptibilities} 
\label{sec:MP}
        
\begin{figure*}[!htb]
\centering
   \includegraphics[width=\linewidth]{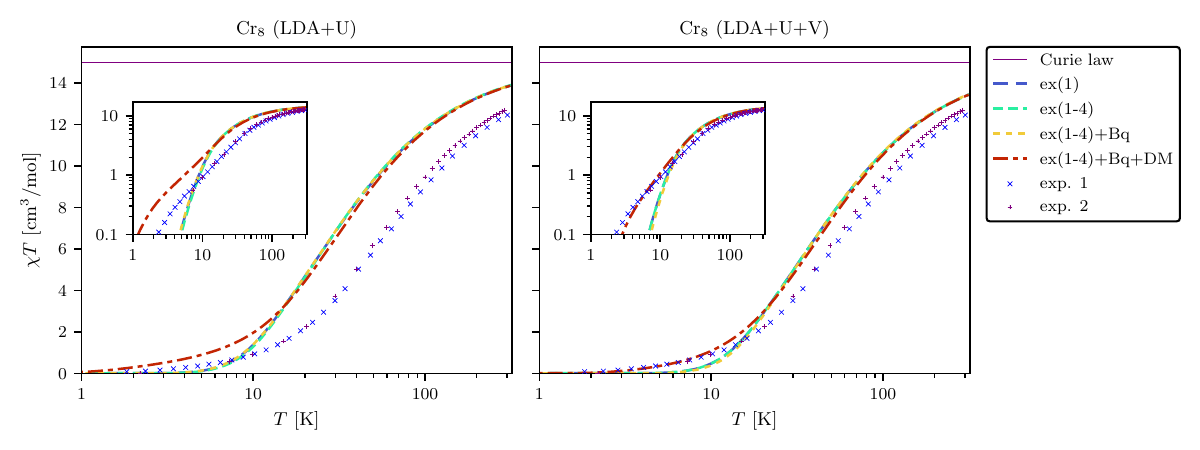}
   \caption{Magnetic susceptibility of \ce{Cr8}. The plots show 
   $\chi$ weighted by the temperature $T$, as obtained from \gls{LDA}+U (left) and \gls{LDA}+U+V (right) by 
   progressively including 
all the computed interactions. The solid purple line corresponds to a system of free spins and therefore the susceptibility recovers the Curie law. The curve \texttt{ex(1)} is obtained from  an anisotropic Heisenberg hamiltonian with only the 1\textsuperscript{st} nearest neighbor interactions, with parameters reported in Table \ref{tab:ncoll-ring}. The curve \texttt{ex(1-4)} includes also the 2\textsuperscript{nd}, 3\textsuperscript{rd} and 4\textsuperscript{th}  nearest neighbor interactions reported in Tables \ref{table:V8-J-fit-non-collinear} and \ref{table:V8-J-fit-non-collinear+V}. The \texttt{ex(1-4)+BQ} curve 
also includes \gls{BQ}s, while \texttt{ex(1-4)+BQ+DM} adds the \gls{DM} interactions, as reported in Tables \ref{tab:biqu} and  \ref{Tab:dm-rings}. 
Blue and purple symbols show the experimental data for \ce{Cr8} from Refs. \cite{van2002magnetic,whinpenny}. 
Insets show the same data with a logarithmic scale on the $y$-axis.}
        \label{fig:sus-Cr8} 
      \end{figure*}
  
      \begin{figure}[!htb]
        \centering
        \includegraphics[width=\linewidth]{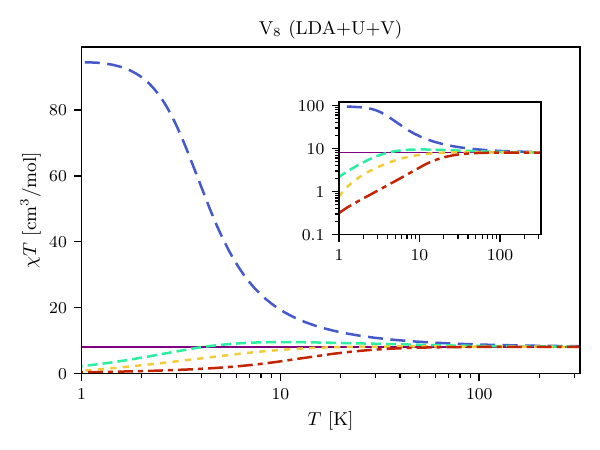} 
        \caption{Magnetic susceptibility of \ce{V8}, with the same legend and considerations of Figure~\ref{fig:sus-Cr8}.} 
        \label{fig:figura-V8}
      \end{figure}
        
We have now determined all the couplings necessary to define the full spin Hamiltonian 
for both \ce{Cr8} and \ce{V8}. This Hamiltonian reads
    \begin{equation}
    \hat H \equiv \hat H_\mathrm{ex} + \hat H_\mathrm{BQ} + \hat H_\mathrm{DM}
      \label{Htot}
    \end{equation}
and corresponds to the sum of Eqs. \ref{hspinbq} and \ref{dmj}. Treating the spins as quantum operators, we proceed to its exact diagonalization that will grant 
access to the full energy spectrum and to the thermal average of any desired operator.
We 
will then 
evaluate 
the magnetic susceptibility, 
i. e., the response of the total magnetization to an applied 
magnetic field. In order to maintain consistency with experiments, we assume that the external magnetic field $B$ is oriented along the $z$ axis. 
With this assumption, the Hamiltonian 
gets augmented by the Zeeman interaction as

      \begin{align}
\hat H\left(B\right) = 
\hat H - \hat{m}_z B
      \end{align}
      where $\hat{m}_z$ is the magnetic moment operator along the $z$ axis, defined as $\hat{m}_z=g\mu_\mathrm{B}\sum_{i}\hat{S}^z_i$. 

      The magnetic susceptibility $\chi$ is defined as the derivative of the total magnetization $M=\braket{\hat{m}_z}_{\beta}$ (i.e. the thermal average of $\hat{m}_z$, $\beta$ is the inverse temperature) with respect to the applied magnetic field $B$: 
      \begin{align}
        \chi = \mu_0 \frac{\partial M }{\partial B } \Bigr\rvert_{B = 0}.
      \end{align}

      where $\mu_0$ is the vacuum permittivity. 
      In systems with localized moments, the modulus is determined by the exchange couplings between electrons on the open $d$ shell of the same \gls{TM} ion. Thus, the total magnetization $M$ results from the degree of alignment of the moments along a certain direction. Ultimately, this alignment depends on the balance between the effective interactions that localized moments exchange with their neighbors 
and the Zeeman coupling to the external field $B$. As a consequence, the susceptibility is expected to be temperature dependent, since temperature changes the number and the thermal population of reachable  states and therefore the capability of the system to vary its magnetization in response to the applied field $B$.
      
      To evaluate the susceptibility $\chi$, we adopt the following procedure:
      \begin{itemize}
        \item the Hamiltonian $\hat{\mathcal{H}}\left(B\right)$  is diagonalized for $B= 0, \pm0.01, \pm0.02$~T;
        \item for each value of $B$, $M$ is evaluated using a canonical distribution for the range of temperatures $T=$0.1-500~K;
        \item for each temperature $T$ the value of $M$ is fitted with a 3\textsuperscript{rd} degree polynomial of the field $B$ which then allows evaluating its derivative at $B=0$ analytically \footnote{For both \ce{Cr8} and \ce{V8} the results are robust with respect to the degree of the adopted polynomial, proving that both the systems can be safely considered in the linear regime within the considered magnetic field intensities.}.
      \end{itemize}

      %
      The results of our calculations are shown in Figs. \ref{fig:sus-Cr8} and  \ref{fig:figura-V8} where the product $\chi T$ is plotted as a function of the absolute temperature $T$ (on a logarithmic scale) for \ce{Cr8} and \ce{V8}, respectively.
      The figures show, in particular, the $\chi T$ curves resulting from the effective Hamiltonians obtained from adding the various kinds of spin-spin interactions one by one, 
so that their contribution can be easily captured. The results are also compared with the Curie limit.

      Fig. \ref{fig:sus-Cr8} reports the results of \ce{Cr8} in comparison with the experimental $\chi T$ data from Refs~\cite{van2002magnetic,whinpenny} within the  temperature range $1-300$~K. 
      Independently of the 
exchange interactions included in the 
spin Hamiltonian, \ce{Cr8} always exhibits the same qualitative behavior -- typical of finite \gls{AFM} systems -- 
compatible with a susceptibility that evolves from vanishing  values at low temperature (where $\chi T \approx 0$) to a 1/T Curie  - like decay  that is rapidly approached at $T \gtrsim 300$ K. 
      Based on the small values of all the effective couplings beyond nearest neighbor exchange (further neighbors exchange and \gls{BQ} in particular) it is not surprising to see all the theoretical curves almost overlapping one another for temperatures exceeding 10 K. At low T, however, some deviations are visible and the curve resulting from the ``full Hamiltonian'' (with also \gls{DM} included) follows the trend of experimental results more closely than simpler approximations, as also highlighted in the insets, 
which show the same results using logarithmic scales on both axes. These results highlight the necessity of including the \gls{DM} interactions in the spin Hamiltonian, even if they are of geometrical origin for this system, in order to recover a consistent representation of exchange couplings  and to gain an accurate description at low T. 

      The comparison between the experimental data and the results obtained from \gls{LDA}+U and \gls{LDA}+U+V shows a clear outperformance of the second corrective scheme over the first. This is especially evident from the inset (log-log) plots showing \gls{LDA}+U+V closely matching the experimental data at low T.  

      Based on the better accuracy of the results obtained for \ce{Cr8} with the extended Hubbard correction, Fig. 10, reporting (in the same range of T) the behavior of $\chi T$ for \ce{V8}, only shows \gls{LDA}+U+V results. Unfortunately no experimental data are available for comparison in this case.
      The most striking aspect  
is the qualitatively different behavior of $\chi T$ when computed with the nearest neighbor exchange couplings only or with more extensive sets of spin-spin interactions. In the first case, since $J_1$ are negative, the system exhibits a \gls{FM} ground state and at sufficiently low temperatures (T $\leq$ $\vert J_1 \vert/K_B \approx$ 10 - 15 K) its susceptibility diverges as $A/T$ causing $\chi T$ to converge to a finite limit \footnote{Note that, while the ground state features the maximum value of the total spin S, the degeneracy of all the states of the multiplet causes the magnetization to vanish.}. The coefficient $A$, proportional to the fluctuation of magnetization around its mean value within the ground state multiplet (at zero field), i.e. $\braket{M^2}_{\beta}-\braket{M}^2_{\beta}$, depends on the total spin 
of the ground state and on the specific types of spin interactions in the Hamiltonian (e.g., whether the system is closer to an Ising or an isotropic Heisenberg model) - see, e.g., Ref. \cite{taylor22}.
At higher T its value converges to the Curie limit, determined by thermal averages over progressively larger sets of states.
      The introduction of second and third nearest neighbor exchange couplings, $J_2$ and $J_3$, featuring a positive sign and a magnitude that is a significant fraction of $J_1$' absolute value (compare Tables \ref{tab:ncoll-ring} and \ref{table:V8-J-fit-non-collinear+V}), stabilizes instead an \gls{AFM} ground state ($S = 0$, non-degenerate) that causes $\chi T$ to rapidly decay towards zero at low temperatures, as in the case of \ce{Cr8}. This is consistent with what is observed in a Heisenberg ring with \gls{FM} nearest neighbor couplings competing with \gls{AFM} interactions between second-nearest neighbors \cite{richter09}. 
      The inclusion of \gls{BQ} and \gls{DM} interactions causes the $\chi T$ curve to decrease faster towards 0 at low temperatures, while the \gls{AFM}-like behavior is qualitatively preserved.

      Obviously, the fact that $\chi T$ tends to zero does not guarantee that $\chi$ vanishes in the same limit. It could in fact be finite or diverge more slowly than $1/T$. More details on these aspects are given in the Appendix \ref{app}, where \gls{FM} isotropic Heisenberg and Ising models are shown to exhibit the same kind of transition to an \gls{AFM} behavior when sufficiently intense next-nearest neighbor \gls{AFM} couplings are turned on.

      In order to complete the  characterization  of \ce{V8} and to better understand, in particular, the nature of the \gls{AFM} state stabilized by a finite $J_2$, we also studied the correlation function between the spins on neighboring sites as a function of  distance and  temperature. 
      \begin{figure}[!t]
        \centering
       \includegraphics[width=\linewidth]{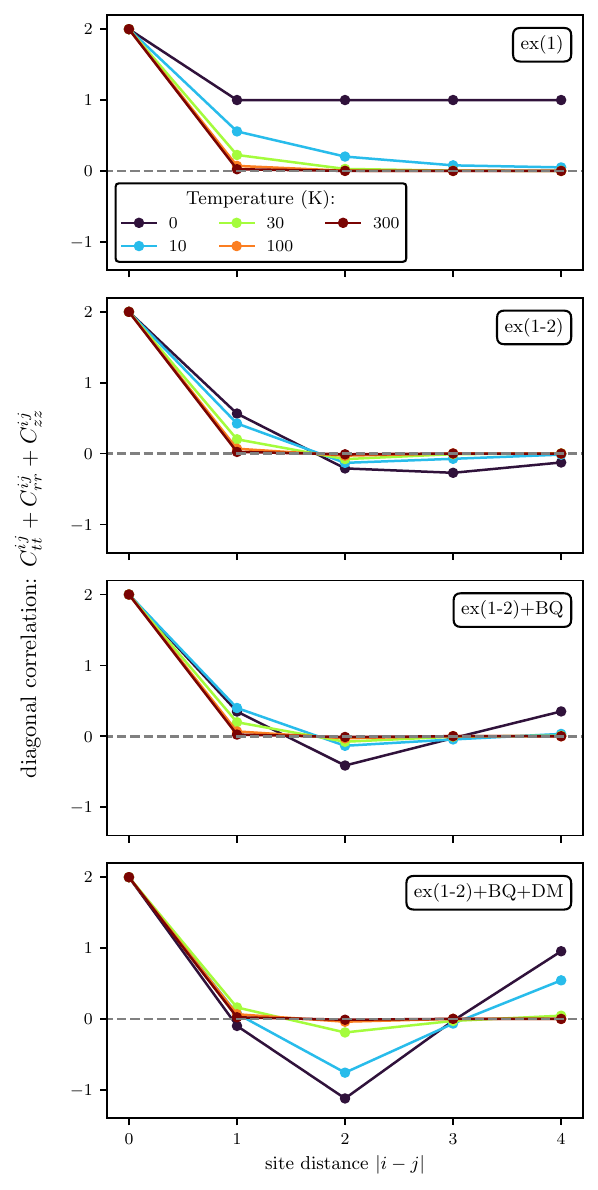}
        \caption{Sum of the diagonal cylindrical components, $C^{ij}_{tt}+C^{ij}_{rr}+C^{ij}_{zz}$, of the spin-spin correlation function of \ce{V8} as a function of the distance between magnetic centers. 
The spin Hamiltonian was diagonalized including the interactions reported in the boxes (top to bottom): only (\gls{FM}) $J_1$; $J_1$ and $J_2$; exchange and \gls{BQ} terms; exchange, \gls{BQ} and \gls{DM} terms. Sites 5 - 7 are omitted for symmetry reasons.}
       \label{corrv8}
      \end{figure}
Fig. \ref{corrv8} compares the diagonal cylindrical spin-spin correlation $C^{ij}_{tt}+C^{ij}_{rr}+C^{ij}_{zz}$
= $\langle S^t_iS^t_j\rangle + \langle S^r_iS^r_j\rangle + \langle S^z_iS^z_j\rangle$ for various sets of couplings: with only \gls{FM} nearest-neighbor $J_1$ (top graph), after including \gls{AFM} $J_2$ (second graph from the top), with exchange and \gls{BQ} interactions (third graph), and with the complete set of all the computed parameters (bottom graph).

It is immediately evident that, upon including the \gls{AFM} second-nearest neighbor coupling, the system develops a tendency to have, besides the \gls{FM} arrangement with nearest neighbor spins (leading to a positive sign of the correlation, whose value is progressively suppressed by the insertion of BQ and DM interactions), an AFM order with the next nearest neighbor, signalled by a negative peak of the correlation function.
This outcome suggests that as magnetic interactions (in particular $J_2$) are progressively switched on, the ground and low-energy states acquire an increasing contribution 
from spin configurations with a very characteristic AFM pattern in which neighboring spins form parallel \textit{pairs}, while successive pairs alternate their orientation along the ring (an up-up-down-down pattern). 
Notably, these configurations are not unique (e.g., the same spin can be FM coupled with the one on either side) and the substantial equivalence of these possible configurations (in a structurally symmetric molecule) might promote a ground state that is more pronouncedly multi-configurational. Quite expectedly, all the correlations rapidly fade to zero upon increasing temperature.

      \section{Discussion and outlook}\label{sec:disc}


A first important outcome of this work is that conventional, nearest neighbors exchange interactions alone are insufficient to describe some molecular magnets, particularly those with open-shell magnetic centers such as \ce{V8}. An adequate spin Hamiltonian must also include BQ and DM interactions, together with exchange couplings that extend beyond nearest neighbors, as shown in Eq.~\eqref{Htot}.

    %
    %
    While the inclusion of \gls{DM} interactions has been relatively common in the literature dealing with non-collinear magnetic systems, including exchange couplings beyond nearest neighbors or \gls{BQ} interactions is quite rare.
    
    Importantly, no a priori assumption was made on the shape of the  magnetic Hamiltonian, except for assuming a diagonal expression. The one shown in Eq. \eqref{Htot} can thus be considered as the minimal model able to capture the energetics of non-collinear spin configurations in the considered systems, as determined from first principles calculations based on Hubbard-corrected DFT.
    %
    The diagonal form of the exchange interactions might seem quite an arbitrary and restrictive ansatz; however our test calculations on the \ce{V8} ring showed that off-diagonal exchange terms (e.g., $J^{rt}$ or $J^{zt}$) are at most one order of magnitude smaller than the ones included in the Hamiltonian and  can thus be safely neglected. Appendix \ref{test} reports the results in tangential configuration as a representative case (see Fig. \ref{fig:V_Jmix}). The same behavior is observed in the radial and axial configurations, where the fitted energy curves with and without off-diagonal exchange interactions are almost perfectly overlapping.
    
 Concerning the behavior of the considered systems, \ce{Cr8} is confirmed to be a ``standard'' Heisenberg magnet with an \gls{AFM} ground state. While exchange couplings along $z$ 
coincide with the ones obtained previously from collinear calculations \cite{stocco2025}, those along the in-plane directions 
are found to be a significant fraction of the ones along $z$. These results suggest that it is the planar circular shape of the molecule that imparts a slight anisotropy to the exchange Hamiltonian. This is also true for \ce{V8} for which, however, the nearest-neighbor exchange couplings exhibit a negative sign, thus favoring a \gls{FM} ground state.
    
    For \ce{Cr8}, the exchange couplings are negligible for second-nearest neighbors and beyond, \gls{BQ} terms are almost vanishing and \gls{DM} interactions retain only a non-zero $z$ component that has a geometrical origin. 
    In contrast, the spin Hamiltonian of \ce{V8} presents a richer variety of magnetic interactions, notably including \gls{AFM} second nearest neighbor exchange couplings (also quite isotropic), relevant \gls{BQ} interactions and significant \gls{DM} couplings. 
    Particularly intriguing from an applicative perspective seems the fact that various magnetic interactions are in competition with one another (e.g., \gls{FM} nearest neighbor vs \gls{AFM} next nearest neighbor exchange), a situation that could take some of the V-based systems on the verge of interesting magnetic transitions. 
    
    A more complete picture of the behavior of both \ce{Cr8} and \ce{V8} was obtained from the exact diagonalization of the spin Hamiltonian and from the evaluation of the thermal averages of some important quantities. 
    
   From the energy spectrum, \ce{Cr8}  
    is confirmed to exhibit an \gls{AFM} ground state, in agreement with  literature. Its overall qualitative behavior is independent of the details of the Hamiltonian and the inclusion of exchange couplings between second-nearest neighbors and of \gls{BQ} interactions has only very marginal effects, also consistently with their small value. The \gls{DM} interactions prove useful to refine the agreement with the experimental trends, especially at low T, below 10-15 K.   Notably, we found a marked improvement of the agreement with experiments, achieved with \gls{DFT}+U+V, compared to \gls{DFT}+U. This highlights the importance of inter-site electronic interactions 
    in determining the value of the magnetic couplings. The higher quality of the computed spin interactions  is also confirmed by the fact that \gls{DFT}+U+V achieves values compatible with the planar geometry of the systems and with their cylindrical symmetry.
    
    As for \ce{V8}, to the best of our knowledge,  this system has never been synthesized nor theoretically considered before. 
    The most striking result is a clear shift from a \gls{FM} to an \gls{AFM} behavior when second nearest-neighbor exchange couplings (of \gls{AFM} character) are inserted in the Hamiltonian, while relatively minor effects are produced by \gls{BQ} and \gls{DM} interactions, in spite of their relatively high values. This radical change of behavior would have been missed if only (\gls{FM}) nearest neighbor interactions were used, confirming the possibility of interesting trends in systems with competing 
    spin-spin interactions.
    Interestingly, the \gls{AFM} ground state of \ce{V8} is found to be qualitatively different from that of \ce{Cr8} in that it is based on alternating pairs of parallel neighboring spins rather than alternating single spins.
    Remarkably, 
    this spin configuration 
    has never emerged as the ground state of \ce{V8}, although its energy gets closer to the \gls{FM} one within \gls{DFT}+U+V.
This result could be the consequence of   
a more pronounced multi-configurational character of the new AFM ground state, in which the alternating pairs of parallel neighbor spins  resemble the double bonds of resonating  Kekulé configurations in aromatic rings.
Thus, the corresponding ground state energy might be intrinsically more difficult to capture with the broken-symmetry configurations accessible through \gls{DFT}.
       
    %
    %
    Other unusual aspects of \ce{V8}, even when calculations do not include atomic SOC, include the emergence of both finite (non geometrical) DM couplings and of significant anisotropies in the magnetic interactions. Assessing the impact of atomic \gls{soc} on these couplings is certainly worth pursuing in a future work. However, as also implied by our results, atomic \gls{soc} alone may be insufficient to  explain the emergence of sizeable \gls{DM} and BQ couplings. The idea that they could be related to the presence of strongly correlated electrons is already present in literature \cite{borisov21} and could be further tested by studying the role of a less-than-half-filled subshell of $d$ states in V cations, as already preliminarily  done in our previous work \cite{stocco2025}.
    Other ideas to explain the origin of DM couplings without atomic \gls{soc} are  gaining momentum in literature.
    %
    Among these, given the ring shape of the molecules,  the hypothesis that finite DMs could rise from ground state'
 spin-currents generating spin-orbit-like interactions (e.g., of Rashba type) \cite{spin-curr,meyer23,Sadamichi,PhysRevLett.116.247201,PhysRevLett.121.147203,cardias2020dzyaloshinskii,Cardias2020,droghetti22,bercioux15} seems particularly attractive and worth further investigation. 

It would be also interesting to extend the study of our systems under dynamical conditions. The  magnon dispersion, directly computed through a recently developed time-dependent \gls{DFT} formalism augmented with Hubbard functionals \cite{Binci2025}, would represent 
    a very important complement to the spectrum of spin excitations determined in this work. 
    
    While exploring these directions is certainly interesting and relevant, it is our impression that this type of calculation, especially for systems containing open-shell atoms, is already pushing the current \gls{xc} energy approximations to the limit of their capabilities. Functionals containing an improved account of non-collinear magnetism and of the effective magnetic fields arising from their \gls{xc} terms \cite{sharma18,persson25}, more general and accurate formulations of the spin-orbit couplings (beyond atomic SOC), and/or an explicit dependence on charge and spin currents \cite{spin-curr,pittalis17,pittalis24,comaskey22,desmarais25}, are envisioned to become an important asset to improve the quantitative accuracy of the type of calculations presented here.


  
  \section*{Acknowledgements}
    The authors are grateful to S. Picozzi, S. Carretta, P. Santini, A. Droghetti, and A. Chiesa for very useful discussions.
    M. B. M. and M. C. acknowledge  support from the Italian National Quantum Science and Technology Institute (PNNR MUR project PE0000023-NQSTI). 
    E. S. acknowledges support from the UFAST International Max Planck Research School.
    L. B. acknowledges the Fellowship from the EPFL QSE Center ``Many-body neural simulations of quantum materials'' (Grant number 10060).
    Calculations of this work were partially realized thanks to the high-performance computing resources and support made available by CINECA (through awards within the ISCRA initiative) and partially on the HPC system ADA at the Max Planck Computing and Data Facility.
    Moreover, via A. F. membership of the UK's HEC Materials Chemistry Consortium, funded by EPSRC (EP/X035859), this work also used the ARCHER2 UK National Supercomputing Service (http://www.archer2.ac.uk). 

  \appendix
  
\section{Additional tests on the \texorpdfstring{V$_8$}{V8} system} 
\label{test}    
\begin{figure}[!b]
\centering
        \includegraphics[width=0.98\linewidth]{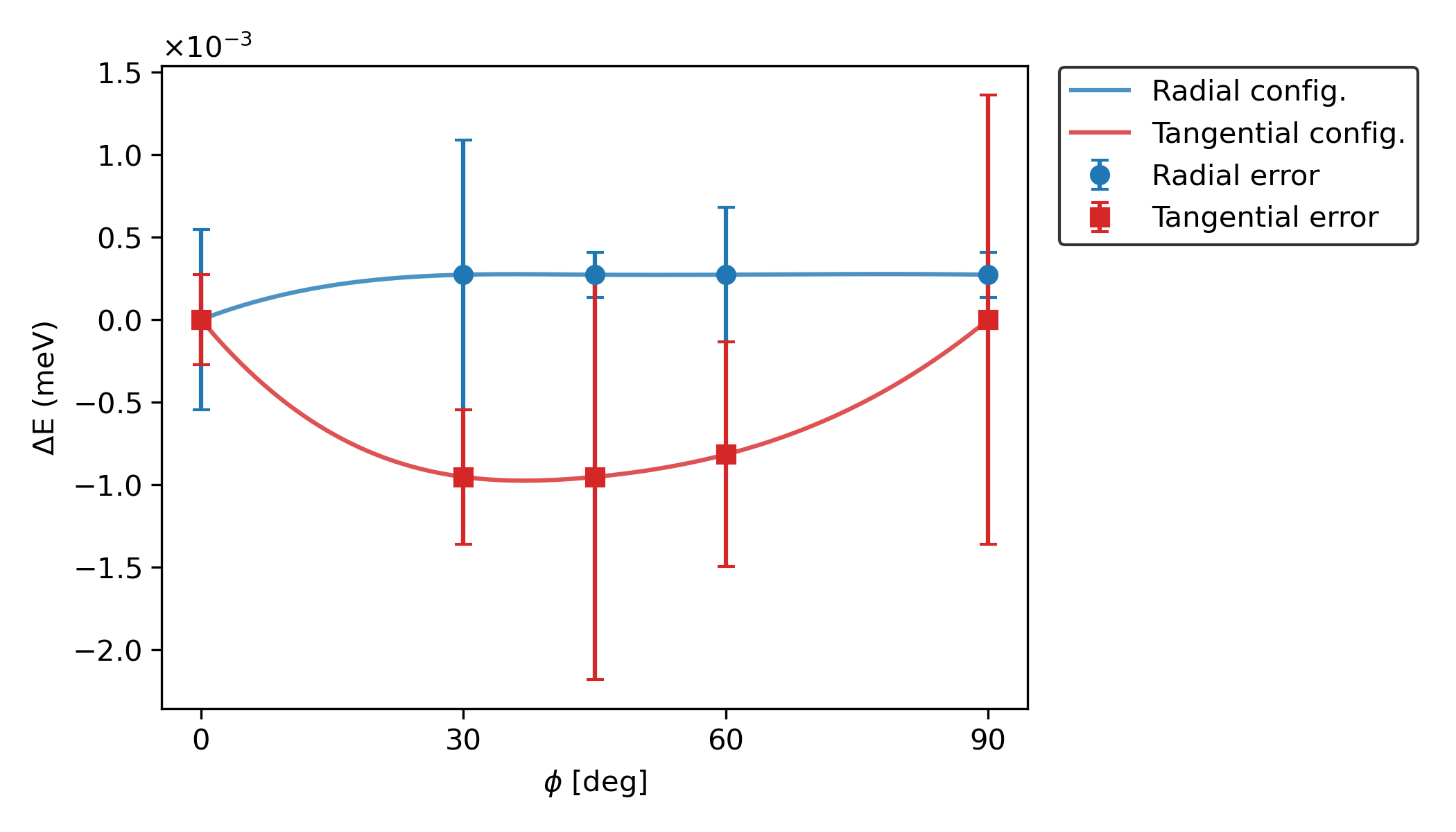}
        \caption{Calculated energy as a function of the spin-rotation angle for the spin localized on the V center of the V\ce{Sc7} ring in both radial and tangential configurations. }
        \label{fig:V+7Sc}
      \end{figure}
Fig.  \ref{fig:V+7Sc} shows the energy profile relative to the rotation of the single V spin (in the $0^\circ$ - $90^\circ$ range) in the V\ce{Sc7} ring. The two curves are obtained for rotations on radial and tangential planes. The very small 
energy variations (about three orders of magnitude smaller than those achieved for the calculation of other interaction parameters) demonstrate that crystal-field-induced anisotropies (typically related to the presence of \gls{soc}) are consistently absent in our systems (see also the discussion in Section \ref{bqt}).

\begin{figure}[!t]
\centering
\includegraphics[width=1.0\linewidth]{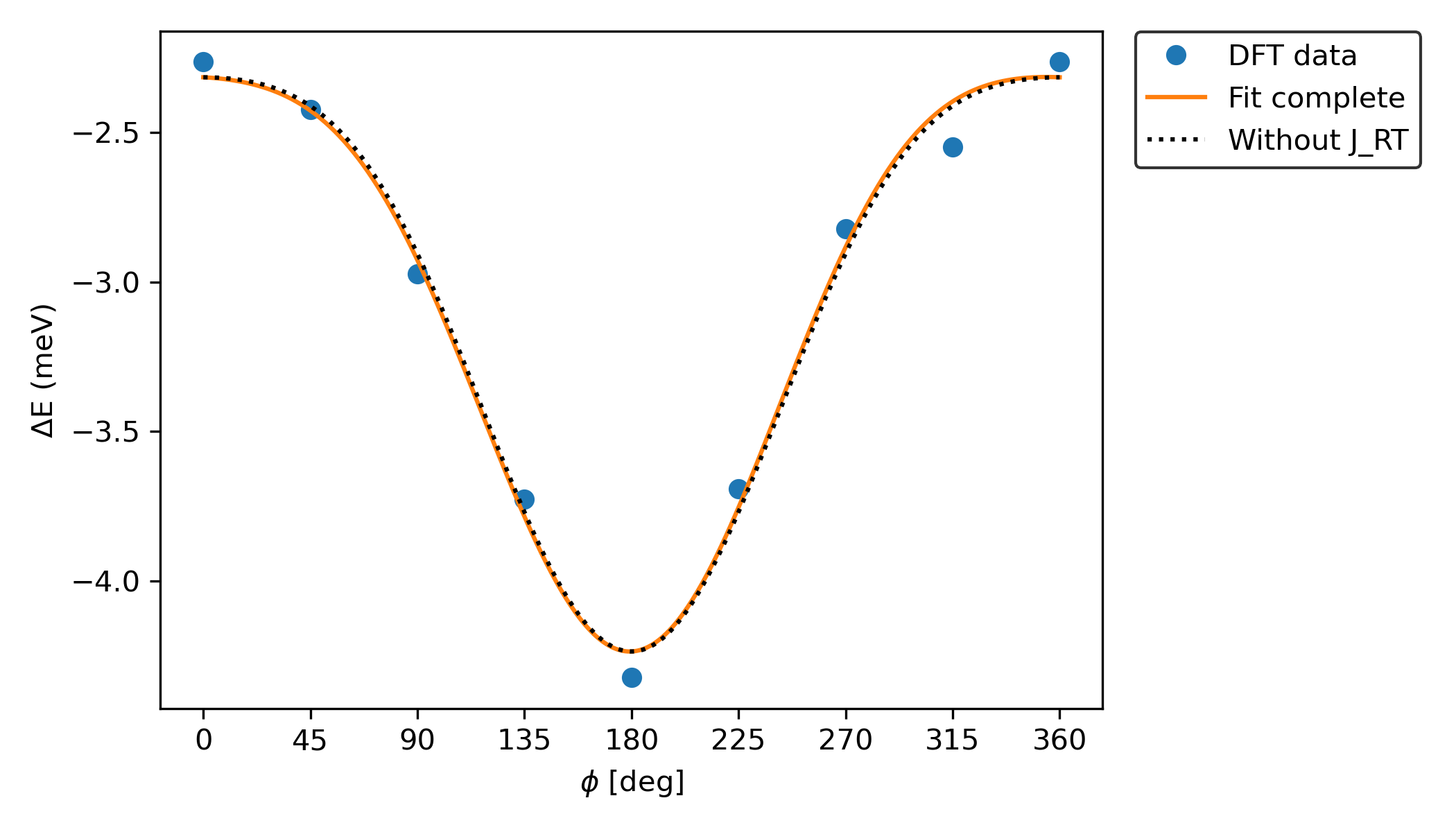}
\caption{Comparison of fitted energy curves with and without off-diagonal exchange terms (tangential configuration) for the \ce{V8} system. A single spin is being rotated from its reference orientation. Data points are the same shown in Fig. \ref{fig:non-collinear-V8} of the main text.}
\label{fig:V_Jmix}
      \end{figure}

Fig.  \ref{fig:V_Jmix} shows the energetics of the rotation of a single spin of \ce{V8} in the tangential spin configuration. The data are the same as those reported in the central graph of Fig. \ref{fig:non-collinear-V8} of the main text. Fits including the off-diagonal exchange components $J^{rt}_{BQ}$ are compared with those restricted to the diagonal terms. The two curves are almost identical. This outcome confirms that off-diagonal interactions are at least one order of magnitude smaller than the others and can be safely neglected for the systems considered here.  
 
\section{Magnetic susceptibility in \texorpdfstring{V$_8$}{V8} and in model Hamiltonians}\label{app}
  
\begin{figure*}
\includegraphics[width=\linewidth]{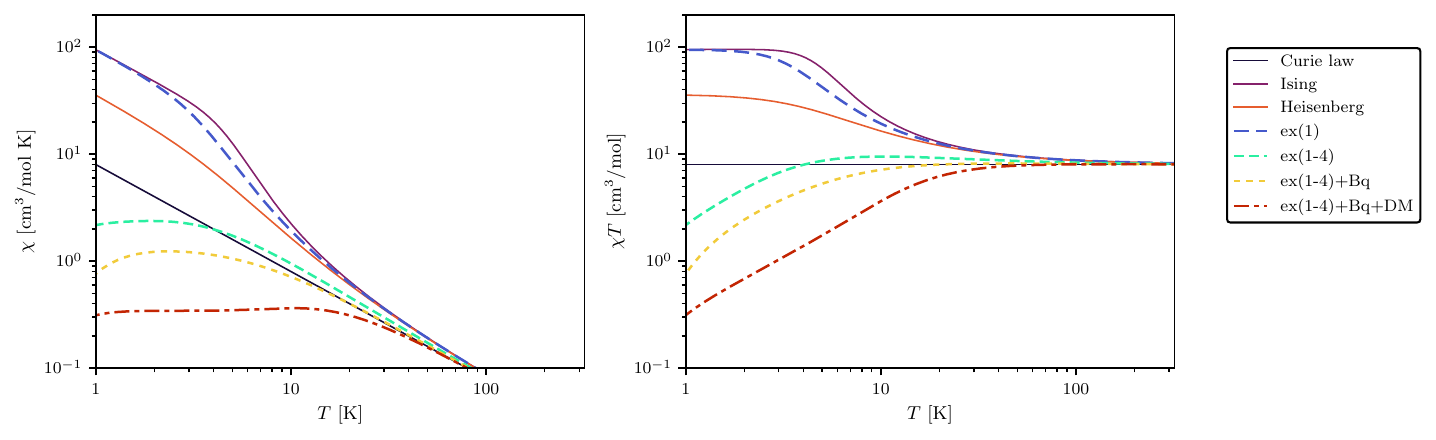}
\caption{$\chi$ (left panel) and $\chi T$ (right panel) as functions of the  temperature $T$ for the same approximations considered in Fig. \ref{fig:figura-V8}. Results from the Heisenberg and Ising models are also included.}
\label{chi}
\end{figure*}
\begin{figure*}
\includegraphics[width=\linewidth]{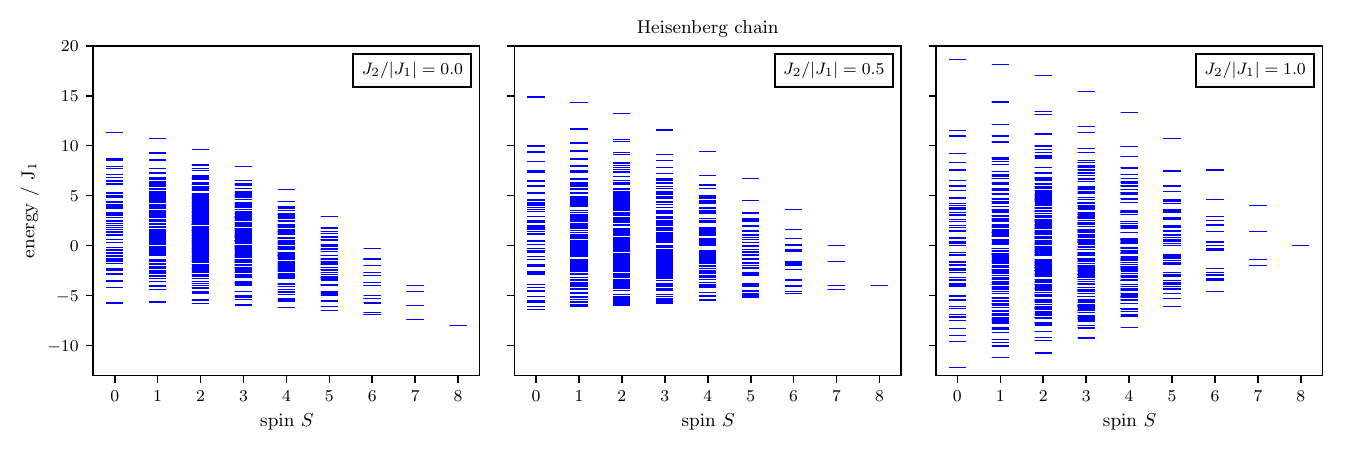}
\includegraphics[width=\linewidth]{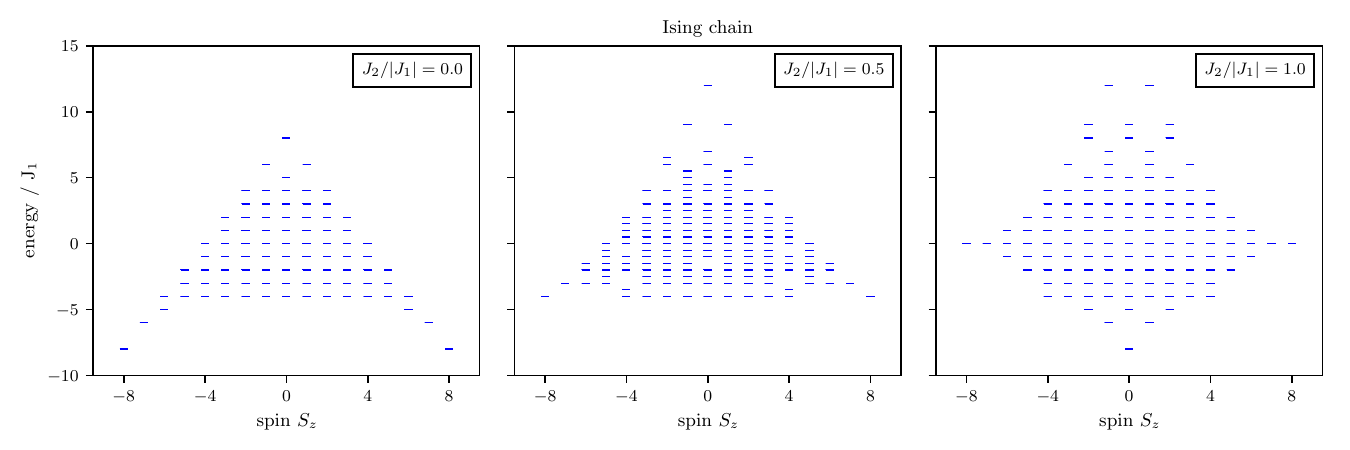}
\caption{Spin-resolved energy spectra for an $N=8$ spin-1 Heisenberg model (upper panels) and an Ising model (lower panels) with \gls{FM} nearest neighbor coupling $J_1=-1\mathrm{meV}$ and next-nearest neighbor interaction $J_2/|J_1| \in [0,1]$. Each panel shows a selected value of the coupling ratio $J_2/|J_1|$, with eigenenergies plotted as horizontal segments positioned at their corresponding total spin quantum number $S$ for the Heisenberg and the total spin $S_z$ for the Ising model. As $J_2/|J_1|$ increases, the low-energy spectrum undergoes a clear restructuring: low-spin states are progressively stabilized relative to high-spin states, reflecting the increasing influence of competing \gls{AFM} next-nearest neighbor interactions. For small $J_2$, the spectrum is dominated by the maximal-spin sector consistent with a \gls{FM} ground state, while larger $J_2$ leads to a redistribution of spectral weight toward lower total spin, indicating a crossover toward an \gls{AFM}-like regime.} 
\label{fig:spectra}
\end{figure*}

This Appendix aims to clarify the magnetic susceptibility of \ce{V8} by discussing the main factors that determine the behaviour of the system described in the main text. To this end, the results presented in Fig. \ref{fig:figura-V8} 
($\chi T$ vs $T$) are reproposed in the log-log graph in the right panel of Fig. \ref{chi}, which also features curves obtained from the diagonalization of Ising and Heisenberg model Hamiltonians containing nearest-neighbor interactions with $J=-0.585$~meV, equal to the coupling parameter $J_1^z$ reported in Table \ref{tab:ncoll-ring}, as obtained from \gls{LDA}+U+V calculations.
      It is easy to realize that, in the limit of low T the \ce{V8} system assumes the behavior of an Ising spin ring (whose susceptibility diverges as 1/$T$) when only nearest neighbor $J$s are taken into consideration. However, as soon as the second-nearest neighbor $J_2$ is switched on, the susceptibility converges to a finite value at low T (that becomes even lower when \gls{BQ} and \gls{DM} interactions are also included) that causes $\chi T$ to tend to zero. 

      The convergence with the behavior of the anisotropic Ising model at low T proves that the latter can be used as a valid proxy system to understand the radical shift in the behavior of \ce{V8} when (\gls{AFM}) $J_2$ are turned on. We then computed the energy spectrum of an 8-site \gls{FM} ring, described either with a Heisenberg or an Ising Hamiltonian ($J_1 =-1$~meV and $J_2 > 0$) for three values of $J_2/\vert J_1\vert$: 0, 0.5 and 1. 
      Fig. \ref{fig:spectra} reports the results obtained.
      The three graphs in each row show, in particular, the energy of the various eigenstates in dependence of the total spin quantum number $S$ for the Heisenberg and the total spin $S_z$ for the Ising model. As it can be easily observed, the spectrum of both these models, although very different from one another, undergoes a significant restructuring when higher values of $J_2$ are considered. For the Heisenberg model, while the energy of high-$S$ states increases, low-$S$ states become progressively more stable and a global singlet ($S = 0$) state becomes the ground state of the system already at $J_2/\vert J_1\vert = 0.5$. The same arguments hold for the Ising model by replacing $S$ with $|S_z|$. This drastic reshaping of the spectrum of the system with the stabilization of low-spin states is what causes the susceptibility to drop at low temperatures, due to a corresponding decrease of the degeneracy of the ground state and the consequent deterioration of the capability of the system to respond to external magnetic fields in terms of variation of its magnetization.
      Comparing the results from Tables \ref{tab:ncoll-ring} and \ref{table:V8-J-fit-non-collinear+V} it is easy to realize that for the \ce{V8} ring considered in this work the $J_2/\vert J_1\vert$ ratio is actually high enough to justify the stabilization of an \gls{AFM} ground state. The anisotropy of the exchange couplings could actually lead to a faster decrease of the ground state degeneracy upon raising $J_2$, causing a stronger tendency of \ce{V8} to assume an \gls{AFM} behavior.

\section{Spin-spin correlations: comparing the \texorpdfstring{\gls{AFM}}{AFM} ground states of \ce{Cr8} and \ce{V8}}  \label{spincorr}

\begin{figure*}
\centering
\includegraphics[width=\linewidth]{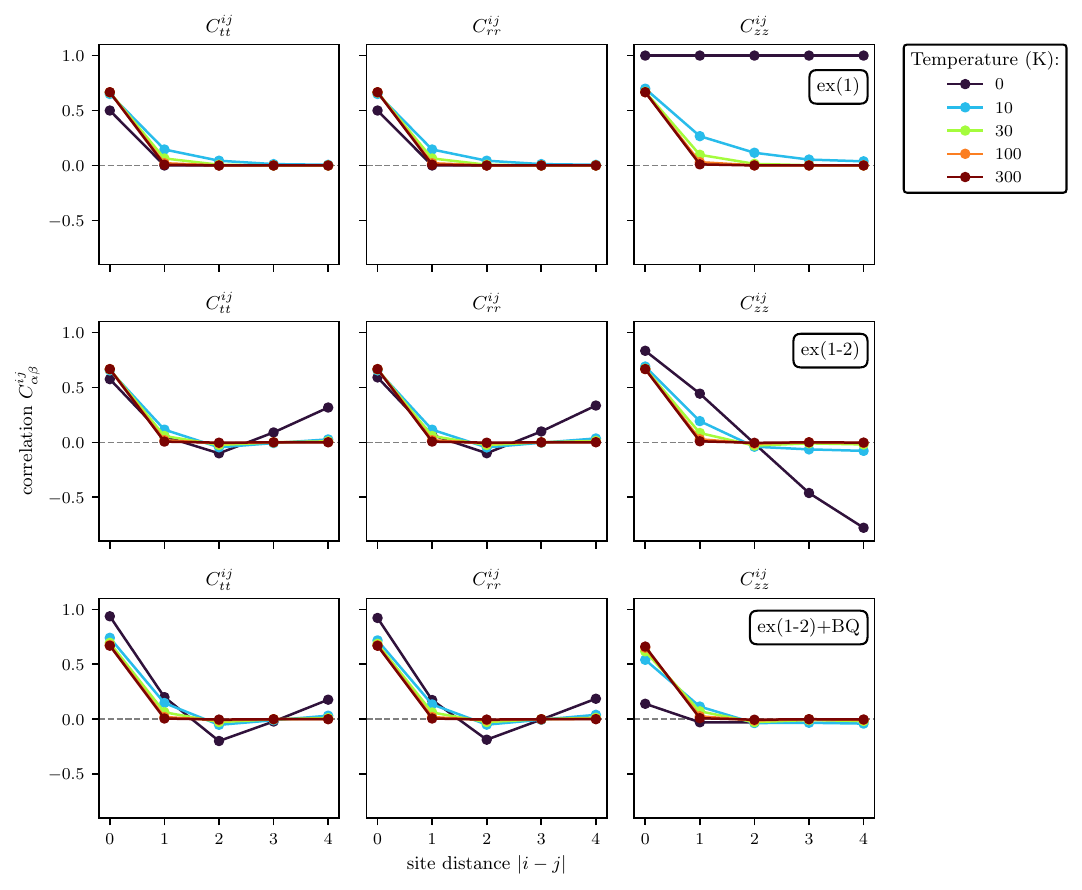}
\caption{Diagonal cylindrical ($tt$, $rr$, and $zz$) components of the spin-spin correlation function as a function of intersite distance, where $t$, $r$, and $z$ denote the tangential, radial, and axial directions, respectively. The color code indicates the absolute temperature, as shown by the bar on the right. The panels show, from top to bottom, the results obtained with nearest-neighbor exchange couplings only, with the inclusion of next-nearest-neighbor $J_2$, and with the additional inclusion of \gls{BQ} couplings. All off-diagonal cylindrical components vanish in these cases.}
\label{chi_j}
\end{figure*}

\begin{figure*}
\centering
\includegraphics[width=\linewidth]{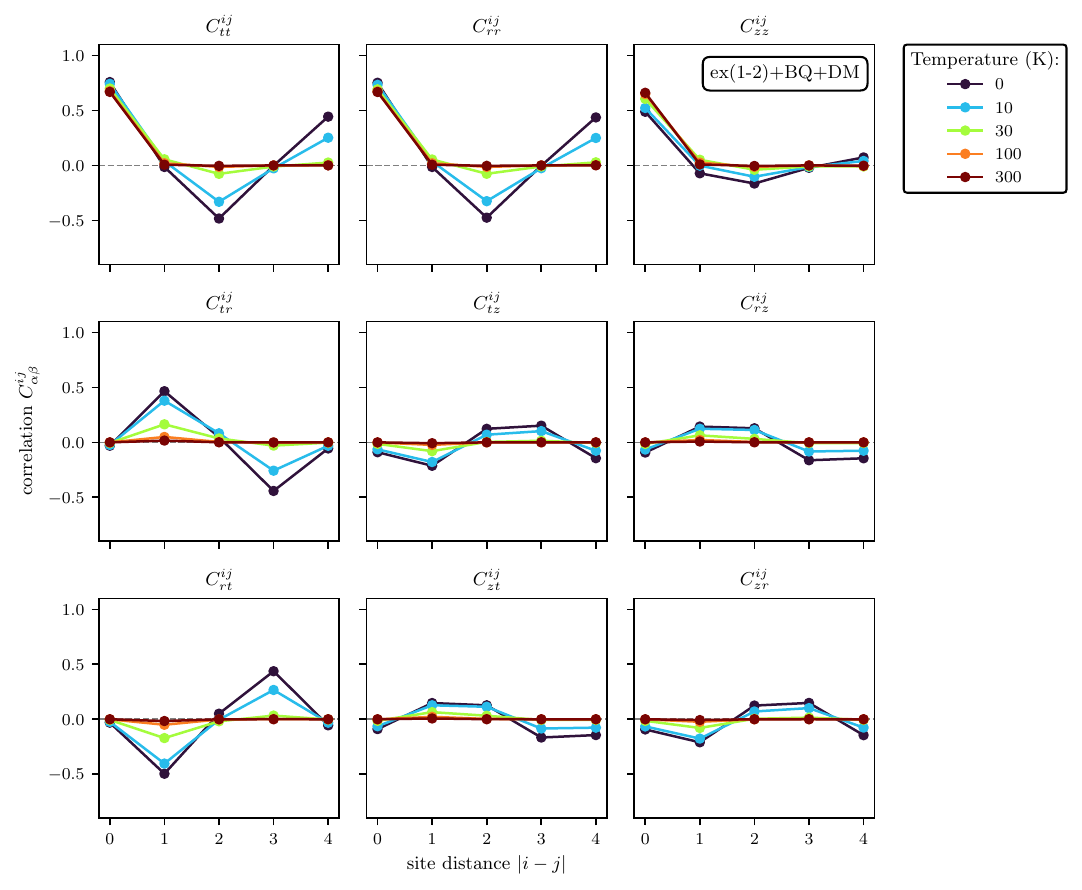}
\caption{All cylindrical components of the spin-spin correlation function as a function of intersite distance after including exchange, \gls{BQ}, and \gls{DM} couplings. Here $t$, $r$, and $z$ denote the tangential, radial, and axial directions, respectively. The color code indicates the absolute temperature, as shown by the bar on the right. The off-diagonal cylindrical components are nonzero in the presence of \gls{DM} interactions.}
\label{chi_bqdm}
\end{figure*}

This Appendix presents the spin-spin correlations resolved in the local cylindrical basis $(t,r,z)$, where $t$ and $r$ denote the tangential and radial in-plane directions, respectively, and $z$ is the axial direction perpendicular to the molecular plane. We define $C^{ij}_{\alpha\beta}=\langle\hat S_i^\alpha\hat S_j^\beta\rangle$, with $\alpha,\beta\in\{t,r,z\}$. The quantity shown in Fig. \ref{corrv8} and discussed in the last part of Section \ref{sec:MP} is the sum of the diagonal cylindrical components, $C^{ij}_{tt}+C^{ij}_{rr}+C^{ij}_{zz}$. Figs. \ref{chi_j} and \ref{chi_bqdm} illustrate the individual components in dependence of the distance between magnetic sites. The color of the different lines indicates the absolute temperature, according to the color bar on the right of each plot. Fig. \ref{chi_j} reports only the diagonal components $C^{ij}_{tt}$, $C^{ij}_{rr}$, and $C^{ij}_{zz}$, all off-diagonal cylindrical components being identically zero without \gls{DM} interactions. 
   From top to bottom, it contains the results obtained with only $J_1$, with both $J_1$ and $J_2$, and after including the \gls{BQ} couplings. Fig. \ref{chi_bqdm} reports all the cylindrical components, notably including also off-diagonal term that are different from zero after the \gls{DM} couplings are included in the Hamiltonian.

It is easy to realize a clear difference between the in-plane components (the almost equivalent $C^{ij}_{tt}$ and $C^{ij}_{rr}$) and the axial component $C^{ij}_{zz}$. In the top panel of Fig. \ref{chi_j}, all diagonal components exhibit a \gls{FM} behavior (with $C^{ij}_{zz}$ being the most pronounced), whereas the middle panel shows a markedly different behavior for the tangential/radial and axial components: the former are slightly negative only for third neighbors (on either side), while the latter shows deep negative values for sites 3--5. The introduction of the \gls{BQ} couplings bottom panel of Fig. \ref{chi_j}) shifts the tangential and radial components to lower values between sites 2 and 6 and almost entirely compensates the negative dip between the same sites introduced by $J_2$, thus leading to a correlation function that, apart from the \gls{FM} neighborhood of the reference site, exhibits an almost flat behavior. The inclusion of \gls{DM} couplings in the Hamiltonian (Fig. \ref{chi_bqdm}) leads to quite significant changes. For the diagonal terms\textcolor{red}{,} one can observe a series of peaks of alternating sign for sites 2 and 4 ( 6 is equivalent to 2 by symmetry), with very small correlations at the intermediate sites (note also the slightly negative value for nearest neighbors in $C^{ij}_{zz}$). At the same time, the off-diagonal cylindrical components become nonzero and comparable to the diagonal ones. While this is in line with the nature of the DM couplings, which favor a helical rotation of the spins around specific axes, it is important to note that reversing the order of the spin components does not give the same correlation function but rather a shifted one (i.e., the correlation matrix is no longer symmetric). Both the amplitude of the oscillation (which is attenuated for off-diagonal correlations involving the $z$ components) and the shift between transposed elements depend on the specific components being considered.
  
    \bibliographystyle{apsrev4-2}
    \bibliography{bibliography}

\end{document}